# An Accelerated Stackelberg Game Approach for Distributed Energy Resource Aggregator participating in Energy and Reserve Markets Considering Security Check

Zhijun Shen[a], Mingbo Liu[a], Lixin Xu[b, *], Wentian Lu[a]

[a] School of Electric Power Engineering, South China University of Technology, Guangzhou 510640, China

[b] Department of Electronic Business, South China University of Technology, Guangzhou 510640, China

**Abstract:** With increasing distributed energy resoures (DERs) integration, the strategic behavior of a DER aggregator in electricity markets will significantly affect the secure operation of the distribution system. In this paper, the interactions among the DER aggregator, energy and reserve markets, and distribution system are investigated through a single-leader-multi-follower Stackelberg game model with the DER aggregator as the leader and the independent system operator and distribution system operator as the followers. To guarantee the operation security of the distribution system, security check problems under three different scenarios are involved in the follower level, which is linearized using a mixed-integer linearized power flow model. Then, using the strong duality theorem, the proposed model is converted into a bi-level mixed-integer linear (BMILP) programming model with only mixed-integer linear follower-level problems. Next, an accelerated relaxation-based bi-level reformulation and decomposition algorithm is proposed to solve the BMILP problem. Finally, case studies are carried out on a constructed integrated transmission and distribution (T&D) system and a practical integrated T&D system to verify the effectiveness of the proposed model and algorithm. The simulation results indicate that the available downward reserve of the DER aggregator will decrease with the security limitation of the distribution system.

## Nomenclature

| | |
|---|---|
| *Abbreviation* | |
| DER | Distributed energy resource |
| CDG | Controllable distributed generator |
| ESS | Energy storage system |
| PV | Photovoltaic |
| DS | Distribution system |
| T&D | Transmission and distribution |
| DSO | Distribution system operator |
| ISO | Independent system operator |
| CB | Capacitor bank |
| SC | Security check |
| BMILP | Bi-level mixed-integer linear programming |
| MILP | Mixed-integer linear programming |
| LP | Linear programming |
| KKT | Karush-Kuhn-Tucker |
| RBRD | Relaxation-based bi-level reformulation and decomposition |
| A-RBRD | Accelerated relaxation-based bi-level reformulation and decomposition |
| *Sets* | |
| $T$ | Set of time periods, $T = [1,2,\ldots,24]$ |
| $N_{CDG}$ | Set of buses with CDG integration |
| $N_{PV}$ | Set of buses with PV integration |
| $N_{ESS}$ | Set of buses with ESS integration |
| $N_G$ | Set of buses with conventional generator integration |
| $N_L$ | Set of lines in transmission system |

* Corresponding author.

E-mail: 179258186@qq.com (Zhijun Shen); epmbliu@scut.edu.cn (Mingbo Liu); xulixin@scut.edu.cn (Lixin Xu); hnlgtiantian@163.com (Wentian Lu).

| | |
|---|---|
| $N_{DS}$ | Set of buses in the DS |
| $\Omega$ | Set of scenarios of SC, where $\Omega = \{NR, UP, DN\}$, NR: normal scheduling scenario, UP: upward-reserve-activated scenario, DN: downward-reserve-activated scenario |
| $N_{CB}$ | Set of buses with CB integration |
| $N_{DL}$ | Set of lines in the DS |
| *Superscripts and subscripts* | |
| $(\cdot)^s$ | Parameters or variables in scenario $s$, $s \in \Omega$ |
| $(\cdot)_{*,t}$ | Parameters or variables at time $t$ |
| *Parameters* | |
| $\boldsymbol{z}^s_{t,c,l}$ | Fixed combinations of follower-level discrete variables of the DSO |
| $\boldsymbol{c}_{UL}$, $\boldsymbol{d}^s_{UL}$ | Coefficient matrices of leader-level objective function |
| $\boldsymbol{d}^s_{LL,t}$ | Coefficient matrix of follower-level objective function of the DSO |
| $\boldsymbol{A}_{UL}, \boldsymbol{b}_{UL}$ | Coefficient matrix and constant term of leader-level constraints |
| $\boldsymbol{A}^s_{LL,t}$, $\boldsymbol{B}^s_{LL,t}$, $\boldsymbol{C}^s_{LL,t}$, $\boldsymbol{b}^s_{LL,t}$ | Coefficient matrices and constant term of follower-level constraints of the DSO |
| $b_{cdgp,i}$, $b^s_{cdgq,i}$ | Coefficients of CDG at bus $i$ |
| $b^s_{cv,i}$, $b^s_{pvq,i}$ | Cost coefficients of PV at bus $i$ |
| $b_{ch,i}$, $b_{dis,i}$ | Cost coefficients of charging and discharging for ESS at bus $i$ |
| $P_{cdg,i,\max}$ | Maximum active power of CDG at bus $i$ |
| $P_{ch,i,\max}$, $P_{dis,i,\max}$ | Maximum charging and discharging active power of ESS at bus $i$ |
| $P_{av,i,t}$ | Available active power of PV at bus $i$ |
| $r_{cdgup,i,\max}$, $r_{cdgdn,i,\max}$ | Maximum upward and downward reserve capacity of CDG at bus $i$ |
| $E_{i,\min}$, $E_{i,\max}$ | Lower and upper bounds of the capacity of ESS at bus $i$ |
| $\eta_{c,i}$, $\eta_{d,i}$ | Charging and discharging efficiency of ESS at bus $i$ |
| $\varepsilon_{PV}$ | Prediction error of PV |
| $R_{up,agg,t}$, $R_{dn,agg,t}$ | Aggregated upward and downward reserve requirements of inflexible DERs of DER aggregator |
| $P_{pv,i,\max}$ | Installed capacity of PV at bus $i$ |
| $O_{g,i}$, $O_{gup,i}$, $O_{gdn,i}$ | Offer energy, upward reserve, and downward reserve price of a conventional generator |
| $O_{cv,agg}$ | Penalty coefficient of energy curtailment of aggregated inflexible supply of DER aggregator |
| $P_{d,i,t}$ | Demand at bus $i$ |
| $R_{up,t}$, $R_{dn,t}$ | Upward and downward reserve requirements |
| $P_{g,i,\max}$ | Maximum energy output of a conventional generator |
| $r_{gup,i,\max}$, $r_{gdn,i,\max}$ | Maximum upward and downward reserve capacity of a conventional generator |
| $G_{l,agg}$, $G_{l,i}$, $D_{l,i}$ | Transfer distribution factors of DER aggregator, generator $i$, and demand $i$ to line $l$ |
| $P_{l,\max}$ | Maximum capacity of line $l$ in transmission system |
| $\tilde{y}$, $\bar{\boldsymbol{Y}}$, $\boldsymbol{Y}$ | Admittance matrices of the DS |
| $\varphi_{cdg,i}$, $\varphi_{pv,i}$ | Maximum power factor angles of CDG and PV at bus $i$ |
| $\Delta Q_{cb,i}$ | Reactive power output of each group of capacitor banks for CB at bus $i$ |
| $Q_{cb,i,\min}$, $Q_{cb,i,\max}$ | Lower and upper bounds of reactive power output of CB at bus $i$ |
| $\Delta k$ | Tap ratio increment per step for substation transformer in the DS |
| $k_{\min}$, $k_{\max}$ | Lower and upper bounds of substation transformer tap ratio in the DS |
| $V_{i,\min}$, $V_{i,\max}$ | Lower and upper bounds of voltage at bus $i$ |
| $I_{ij,\max}$ | Upper bound of the $ij$-th line current in the DS |
| $g_{ij}$, $b_{ij}$ | Conductance and susceptance of line $ij$ in the DS |
| *Leader-level variables of the DER aggregator* | |
| $\boldsymbol{x}$ | Leader-level variables |
| $P_{cdg,i,t}$ | Active power outputs of CDG at bus $i$ |
| $P^s_{cv,i,t}$ | Curtailed active power of PV at bus $i$ |
| $P_{ch,i,t}$, $P_{dis,i,t}$ | Charging and discharging active power of ESS at bus *i* |
| $\underline{P}_{flex,agg,t}$, $\overline{P}_{flex,agg,t}$ | Lower and upper bounds of flexible energy bidding quantity of DER aggregator |
| $P_{infl,agg,t}$ | Inflexible energy bidding quantity of DER aggregator |
| $o_{flex,agg,t}$ | Energy bidding price of flexible supply of DER aggregator |
| $\overline{r}_{up,agg,t}$, $\overline{r}_{dn,agg,t}$ | Upward and downward reserve bidding quantities of DER aggregator |
| $o_{up,agg,t}$, $o_{dn,agg,t}$ | Upward and downward reserve bidding price of DER aggregator |
| $r_{cdgup,i,t}$, $r_{cdgdn,i,t}$ | Scheduled upward and downward reserve of CDG at bus *i* |
| $r_{essup,i,t}$, $r_{essdn,i,t}$ | Scheduled upward and downward reserve of ESS at bus *i* |
| $I_{ch,i,t}$, $I_{dis,i,t}$ | State of charging and discharging of ESS at bus *i* |
| $E_{i,t}$ | Energy capacity of ESS at bus *i* |
| *Follower-level variables of the ISO* | |
| $P_{flex,agg,t}$ | Scheduled energy of aggregated flexible supply of DER aggregator |
| $P_{cv,agg,t}$ | Scheduled energy curtailment of aggregated inflexible supply of DER aggregator |
| $r_{up,agg,t}$, $r_{dn,agg,t}$ | Scheduled upward and downward reserve of DER aggregator |
| $P_{g,i,t}$ | Scheduled energy of conventional generator at bus *i* |
| $r_{gup,i,t}$, $r_{gdn,i,t}$ | Scheduled upward and downward reserve of conventional generator at bus *i* |
| $\lambda_{*,t}$ | Clearing price |
| $\underline{\nu}_{*,t}$, $\overline{\nu}_{*,t}$ | Dual variables corresponding to inequality constraints of ISO |
| *Follower-level variables of the DSO* | |
| $\boldsymbol{m}_t^s$, $\boldsymbol{z}_t^s$ | Follower-level continuous and discrete variables of the DSO |
| $\boldsymbol{m}_{t,c,l}^s$ | Follower-level added continuous variables of the DSO |
| $Q_{cdg,i,t}^s$ | Reactive power output of CDG at bus *i* |
| $Q_{cv,i,t}^s$ | Reactive power output of PV at bus *i* |
| $V_{i,t}^s$, $\theta_{i,t}^s$ | Magnitude and angle of voltage at bus *i* in the DS |
| $P_{i,t}^s$, $Q_{i,t}^s$ | Net active and reactive power injection at bus *i* in the DS |
| $C_{cb,i,t}^s$ | Number of groups of capacitor banks in operation for CB at bus *i* in the DS |
| $k_t^s$ | Substation transformer tap ratio in the DS |
| $I_{ij,t}^s$ | Current of line *ij* in the DS |

# 1 Introduction

*1.1 Background*

In recent years, the rapid development of distributed energy resources (DERs), including controllable distributed generators (CDGs), renewable energy resources, and energy storage systems (ESSs), has drawn attention across the world [1]. Many countries and regions have set up incentive policies to promote the construction of DERs to accommodate the increasing renewable energy resources, such as the state of California [2]. Meanwhile, the development of the electricity market open environment has created opportunities to expand the profits of DERs [3]. However, the capacity of a single DER is insufficient to participate in the electricity market with market entry barriers. To overcome this problem, the DER aggregator is introduced as an entity to participate in the market through DER aggregation. Neverthless, existing electricity markets, especially day-ahead energy and reserve markets, mainly focus on the active power transactions in the transmission system, but most DERs emerge in a distribution system (DS). The clearing active power scheduling scheme of DERs will significantly affect the operation security of a DS, and even cause voltage problems or congestion problems. Therefore, investigating the interactions among DERs, electricity markets, and the DS have become increasingly important.

*1.2 Literature survey*

The interaction between the DER aggregator and electricity market has been widely investigated over the past few years. Iria *et al.* [4] proposed a two-stage stochastic model to define the bids of an aggregator of prosumers in day-ahead energy market. In [5], the model was extended to further consider day-ahead secondary reserve markets. A decision-making strategy for the renewable energy resource aggregator in day-ahead electricity market was implemented as a two-stage stochastic model based on the conditional value-at-risk method in [6]. Abapour *et al.* [7] investigated the robust participation strategies of the demand response aggregator in energy markets based on game theory. Sarker *et al.* [8] investigated the bidding strategy of the electric vehicle aggregator in energy and reserve markets. In [9], Abbasi *et al.* proposed a three-stage stochastic model for a virtual power plant to participate in a pool-based energy market. The risk-based participation strategy of a power aggregator of commercial compressed air energy storage and wind power in three markets was presented as a three-stage stochastic model in [10]. In [11], Somma *et al.* proposed a stochastic model to investigate the interaction between a DER aggregator and day-ahead energy market, considering the flexibility of demand-side resources. In the studies mentioned above [4-11], DER aggregators were regarded as price-takers, and DERs were scheduled according to the forecast prices of the markets. For those scenarios where DER aggregators were regarded as price-makers, existing research exploited a Stackelberg game framework to analyze the interaction between the aggregators and the electricity market clearing process. The bidding and scheduling problem of the aggregator and the clearing process of the electricity market were respectively modeled as leader- and follower-level problems. Kardakos *et al.* [12] formulated a three-stage stochastic bi-level model for a commercial virtual power plant to participate in day-ahead energy market, and the Stackelberg game framework between the flexible load aggregator and day-ahead energy and reserve markets considering various types of flexible loads was studied in [13]. Xiao *et al.* [14] studied the strategic bidding of the aggregator of prosumers in joint energy and regulation markets by solving a bi-level stochastic model. In [15], a Stackelberg game model was proposed to research the optimal participation of a DER aggregator in day-ahead energy market, accounting for the interactions between DER aggregator and DERs. Bahramara *et al.* [16] investigated the interaction between a price-maker distribution company and day-ahead energy and reserve markets based on Stackelberg game theory. Sheikhahmadi *et al.* [17] further considered the uncertainty of DERs and the risk management in the bi-level decision-making model of the distribution company in day-ahead energy market.

The studies described above [4-17] mainly focused on the interaction between DER aggregators and electricity market. The impacts and limitations of the DS were not considered. This may lead to an over-optimistic estimation of the available services of DERs, and may even cause operation security problem in a DS. Therefore, some researchers further incorporated the network constraints of DS. Haghighat *et al.* [18] studied the interaction between a distribution company and day-ahead energy market, involving the nonlinear power flow model. Zhang *et al.* [19] applied the nonlinear power flow model of a DS to formulate the distribution company trading in day-ahead and real-time markets. However, the nonlinearity of the power flow model greatly increased the complexity of the Stackelberg game problems. Cadre *et al.* [20] applied the convex second-order cone programming relaxation form of the power flow model to formulate the Stackelberg game between the operation problem of a DS and the day-ahead energy market. Bahramara *et al.* [21] presented a single-leader-multi-follower Stackelberg game model for the distribution company’s interaction with day-ahead energy market and microgrids, exploiting the piecewise linearized active power flow equations of a DS. Furthermore, Sheikhahmadi *et al.* [22] applied a similar method to study the Stackelberg game among day-ahead energy market, distribution company, and DERs. However, the reactive power flow of the DS was neglected, which would lead to large approximation errors.

Although some studies, such as [18-22], indicated that a distribution company can simultaneously act as the DER aggregator and consider the limitation of network constraints of the DS, the distribution companies in some countries, such as China, are not independent entities from the distribution system operator (DSO), which is a subsidiary utility of the independent system operator (ISO). Thus, those distribution companies cannot act as participants in electricity markets. However, little research has been done on the game model among the DER aggregator, day-ahead energy and reserve markets, and the DS, especially when considering the impact of the discrete control devices in the DS, such as the substation transformer and capacitor banks (CBs).

### *1.3 Contribution and organization*

In this study, we investigate a novel Stackelberg game model for the strategic behavior of a DER aggregator in day-ahead energy and reserve markets, considering the security check (SC) by the DSO. The interactions among a DER aggregator, day-ahead energy and reserve markets, and the DS are illustrated in Fig. 1. Specifically, the DER aggregator submits the aggregated bid and offer to the ISO for the energy and reserve markets. Next, the ISO clears the market and delivers the clearing prices and scheduled quantities to the DER aggregator, and then the DER aggregator dispatches the DERs in the DS. Simultaneously, the DER aggregator sends the scheduling scheme of DERs to the DSO for the SC in order to guarantee the operation security of the DS. Finally, based on the results of SC of the DS, the DSO delivers the scheduled schemes of DERs, substation transformer tap, and CBs. The contributions of this paper are summarized as follows:

1) A single-leader-multi-follower Stackelberg game model is established to capture the interactions among a DER aggregator, day-ahead energy and reserve markets, and DS, with the DER aggregator as the leader and the ISO and DSO as the followers. This is different from the investigations in previous studies so far, as the SC of the scheduling scheme of DERs by the DSO under three different scenarios is involved in the follower level.

2) The linearized power flow model of the DS in [23] is extended to further consider the substation transformer tap based on the linearized method in [24]. Consequently, the SC model of the DS can be simplified as a mixed-integer linear programming (MILP) model.

3) Based on the relaxation-based bi-level reformulation and decomposition (RBRD) algorithm [25], an accelerated RBRD (A-RBRD) algorithm is developed to solve the proposed Stackelberg game model. Based on the characteristic of the SC problems, the proposed algorithm identifies and eliminates inactive inequality constraints of the added optimality cut and thus reduces the binary variables of the master problem at each iteration by solving some linear programming (LP) problems.

The remainder of this paper is organized as follows. Section 2 depicts the proposed Stackelberg game model. The proposed A-RBRD algorithm for the Stackelberg game model is introduced in Section 3. Section 4 presents case studies, results, and discussions. Finally, Section 5 draws conclusions from this study.

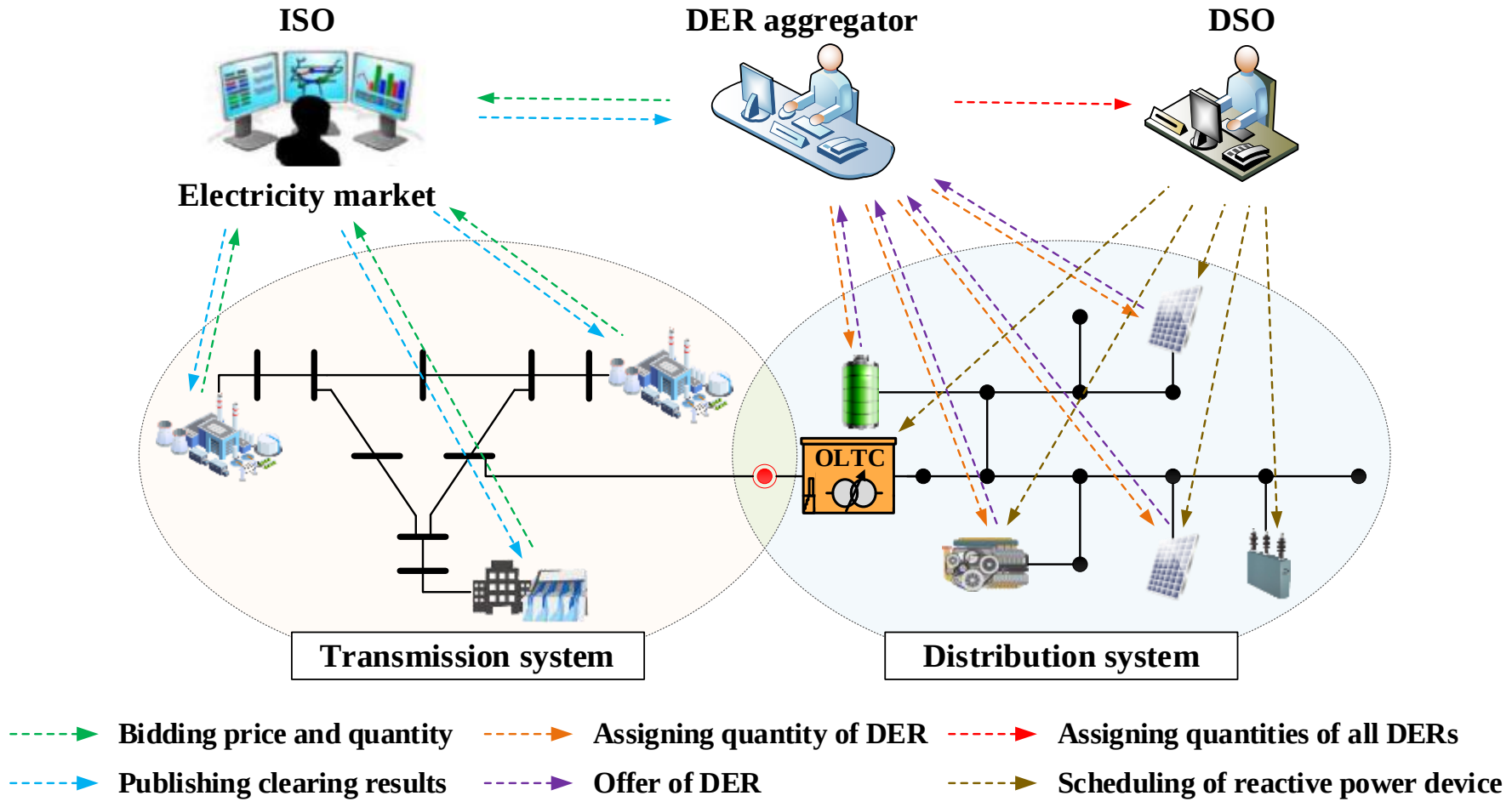

**Fig. 1.** Interactions among the DER aggregator, day-ahead energy and reserve markets, and DS.

## 2 Stackelberg game model for DER aggregator in energy and reserve markets considering SC

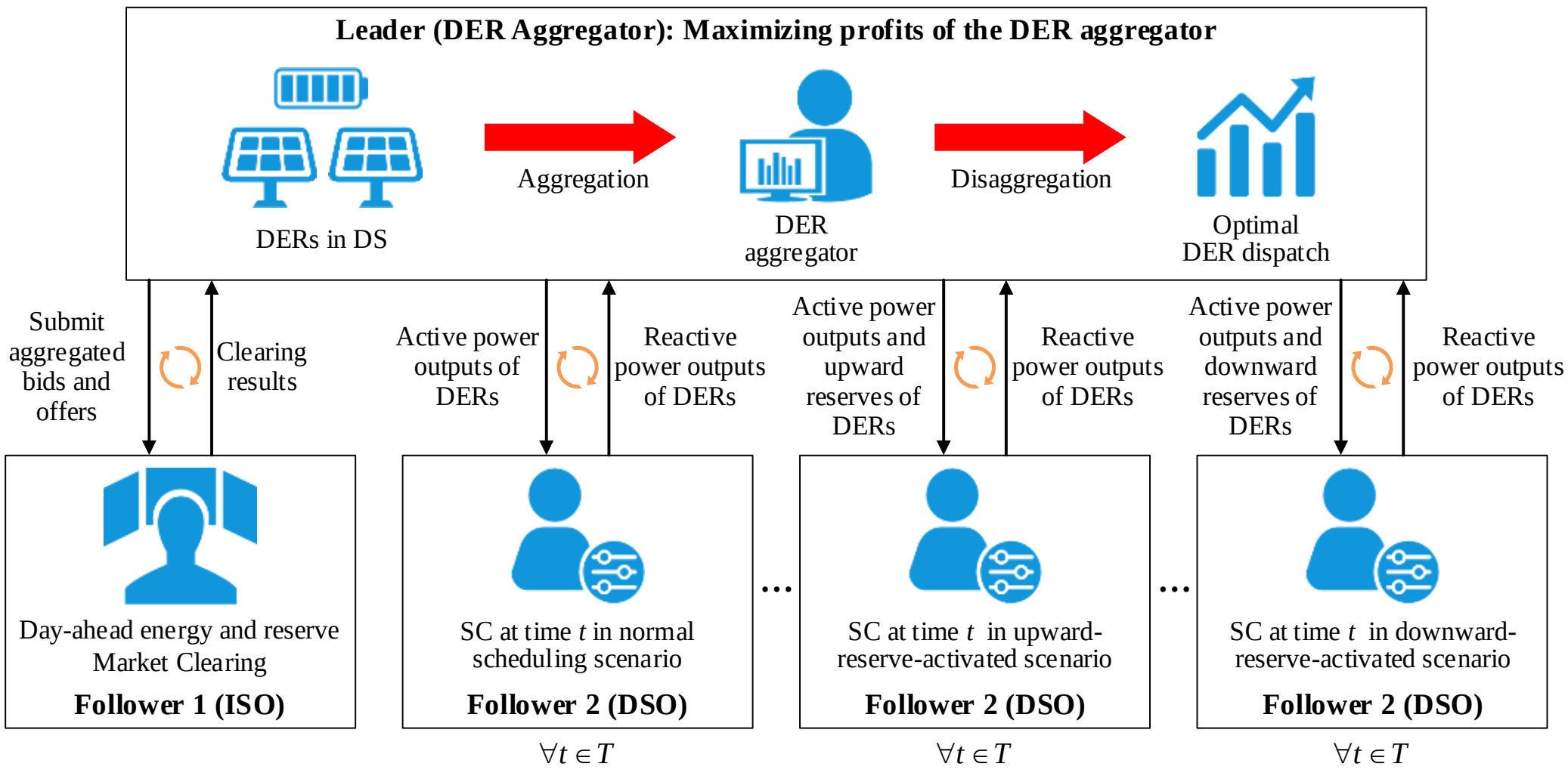

**Fig. 2.** Hierarchical structure of the proposed single-leader-multi-follower Stackelberg game model.

In this section, the interactions among the DER aggregator, day-ahead energy reserve markets, and DS are formulated as a single-leader-multi-follower Stackelberg game model. The hierarchical structure of the proposed model is illustrated in Fig. 2. In the leader-level problem, the DER aggregator optimizes its bidding and scheduling strategy in day-ahead energy and reserve markets considering the SC by the DSO. The effect of the decisions of the DER aggregator on the day-ahead market by the ISO is modeled as a follower-level problem. Moreover, the SC problems of the DS involving discrete control devices in the normal scheduling and reserve-activated scenarios by the DSO are also modeled as follower-level problems.

The assumptions of the proposed model are as follows:

1) The DS is connected to the transmission system through only one substation. The network losses of the DS are neglected.

2) The reactive power exchanged between the transmission and distribution (T&D) systems can be balanced at the boundary substation at each time period.

3) Only one DER aggregator is considered, and all of the DERs in the DS contract with this DER aggregator. The DERs are obliged to provide the reactive power compensation services for the DS within its capacity.

4) Only three scenarios for the SC are considered: i) normal scheduling scenario, ii) upward-reserve-activated scenario, and iii) downward-reserve-activated scenario. It should be noted that the upward-reserve-activated (or downward-reserve-activated) scenario is modeled as the extreme scenario in which the maximal scheduled upward (or downward) reserves are activated when the maximal downward (or upward) reserve requirements of the inflexible DERs are needed.

*2.1 Leader-level problem: Bidding and scheduling model for DER aggregator*

*2.1.1 Objective*

The objective of the leader-level problem is to maximize the profits of the DER aggregator from the day-ahead electricity market, that is,

$$\max \quad F_1 - F_2 - F_3 - F_4 , \tag{1a}$$

$$F_1 = \sum_{t\in T}\Big(\lambda_{en,agg,t}\left(P_{flex,agg,t} + P_{infl,agg,t} - P_{cv,agg,t}\right) + \lambda_{up,t} r_{up,agg,t} + \lambda_{dn,t} r_{dn,agg,t}\Big), \tag{1b}$$

$$F_2 = \sum_{t\in T}\sum_{i\in N_{CDG}}\left(b_{cdgp,i}P_{cdg,i,t} + \sum_{s\in\Omega} b^{s}_{cdgq,i}\left|Q^{s}_{cdg,i,t}\right|\right), \tag{1c}$$

$$F_3 = \sum_{t\in T}\sum_{i\in N_{PV}}\sum_{s\in\Omega}\left(b^{s}_{cv,i}P^{s}_{cv,i,t} + b^{s}_{pvq,i}\left|Q^{s}_{pv,i,t}\right|\right), \tag{1d}$$

$$F_4 = \sum_{t\in T}\sum_{i\in N_{ESS}}\left(b_{ch,i}P_{ch,i,t} + b_{dis,i}P_{dis,i,t}\right), \tag{1e}$$

where $F_1$ represents the revenue from day-ahead energy and reserve markets; $F_2$ denotes the generation costs of CDGs; $F_3$ denotes the penalty costs of the curtailed active power and the generation costs of reactive power of the photovoltaics (PVs); and $F_4$ represents the costs of the charging and discharging of ESSs. Note that the absolute terms in the objective function can be easily linearized by introducing auxiliary variables and inequality constraints [24].

*2.1.2 Constraints*

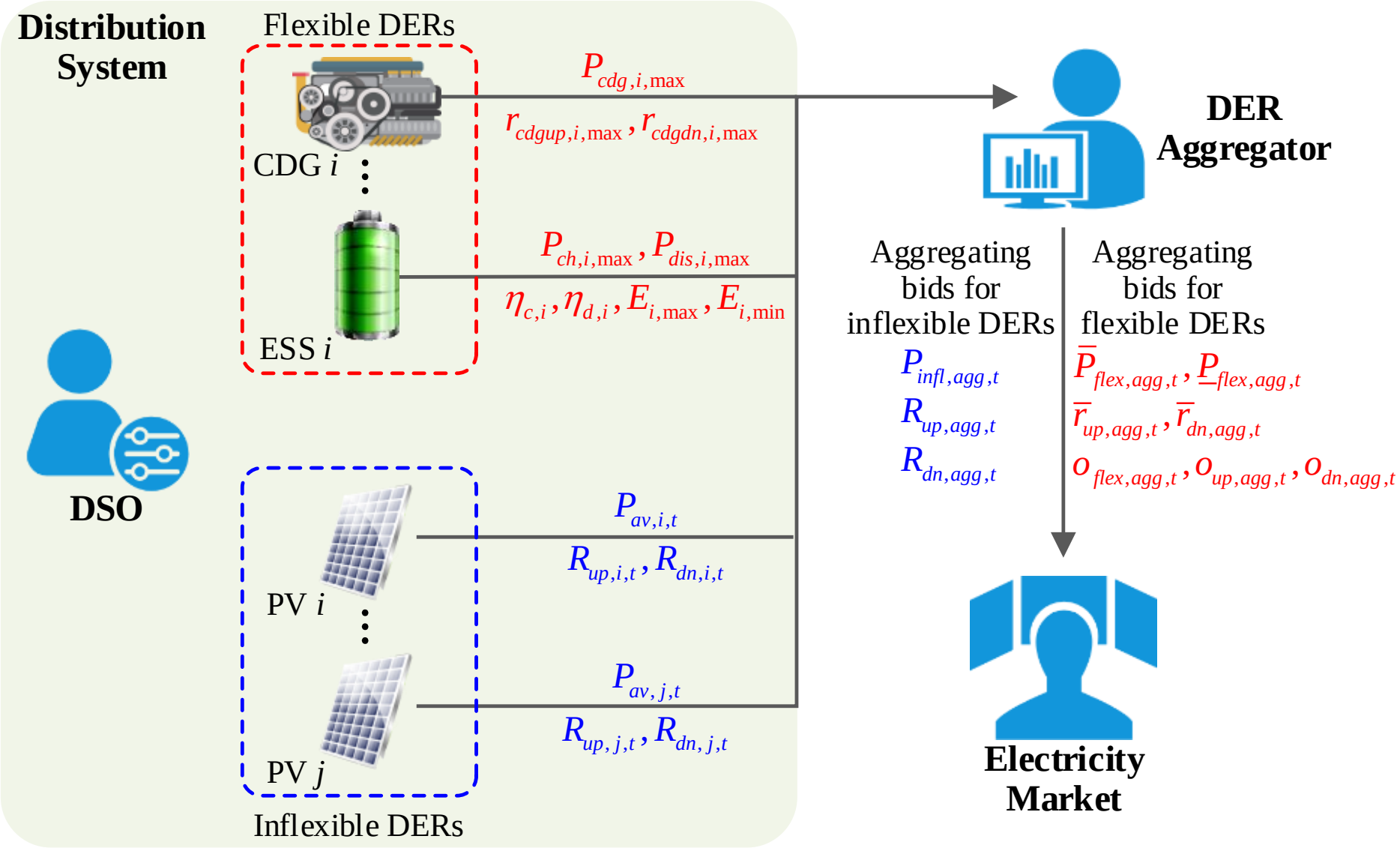

**Fig. 3.** Framework of energy and reserve bids and offers of the DER aggregator.

In this paper, the DER aggregator separately aggregates the flexible and inflexible DERs in the DS into a flexible supplier and an inflexible supplier, as shown in Fig. 3. Therefore, the DER aggregator respectively submits the bids and offers of the aggregated flexible supplier and the available energy quantity and reserve requirements of the aggregated inflexible supplier in energy and reserve markets. The constraints of the energy and reserve bids of the DER aggregator are expressed as follows.

1) Energy bidding constraints of the DER aggregator

$$\bar{P}_{flex,agg,t} \le \sum_{i\in N_{CDG}} P_{cdg,i,\max} + \sum_{i\in N_{ESS}} P_{dis,i,\max} I_{dis,i,t} \,, \tag{2a}$$

$$\underline{P}_{flex,agg,t} \ge -\sum_{i\in N_{ESS}} P_{ch,i,\max} I_{ch,i,t} \,, \tag{2b}$$

$$P_{infl,agg,t} \le \sum_{i\in N_{PV}} P_{av,i,t} \,, \tag{2c}$$

$$o_{flex,agg,t} \ge 0 \,. \tag{2d}$$

Eqs. (2a) and (2b) restrict the aggregated flexible energy bidding quantity. Eq. (2c) defines the aggregated inflexible energy bidding quantity. Eq. (2d) ensures the non-negativity of the energy bidding price for the aggregated flexible energy quantity.

2) Reserve bidding constraints of the DER aggregator

$$0 \le \bar{r}_{up,agg,t} \le \sum_{i\in N_{CDG}} \min\left\{P_{cdg,i,\max} - P_{cdg,i,t}, r_{cdgup,i,\max}\right\} + \sum_{i\in N_{ESS}} \min\left\{P_{dis,i,\max} - P_{dis,i,t} + P_{ch,i,t}, \eta_{d,i}\left(E_{i,t-1} - E_{i,\min}\right)\right\}, \tag{3a}$$

$$0 \le \bar{r}_{dn,agg,t} \le \sum_{i\in N_{CDG}} \min\left\{P_{cdg,i,t} - P_{cdg,i,\min}, r_{cdgdn,i,\max}\right\} + \sum_{i\in N_{ESS}} \min\left\{P_{ch,i,\max} + P_{dis,i,t} - P_{ch,i,t}, \frac{E_{i,\max} - E_{i,t-1}}{\eta_{c,i}}\right\}, \tag{3b}$$

$$R_{up,agg,t} = \sum_{i\in N_{PV}} R_{up,i,t} = \sum_{i\in N_{PV}} \varepsilon_{PV} P_{av,i,t} \,, \tag{3c}$$

$$R_{dn,agg,t} = \sum_{i\in N_{PV}} R_{dn,i,t} = \sum_{i\in N_{PV}} \min\left\{P_{pv,i,\max} - P_{av,i,t}, \varepsilon_{PV} P_{av,i,t}\right\}, \tag{3d}$$

$$o_{up,agg,t} \ge 0, \quad o_{dn,agg,t} \ge 0 \,. \tag{3e}$$

Eqs. (3a) and (3b) restrict the aggregated reserve bidding quantity offered by the flexible DERs. Eqs. (3c) and (3d) define the aggregated reserve requirements of the inflexible DERs. Eq. (3e) ensures the non-negativity of the reserve bidding prices.

Based on the bidding of the DER aggregator, the ISO clears day-ahead energy and reserve markets and delivers the clearing prices and the scheduled quantities back to the DER aggregator. After that, the DER aggregator assigns the scheduled quantities to each DER and sends the allocations of DERs to the DSO for the SC, as depicted in Fig. 4. The relevant constraints of the scheduled energy and reserve of the DERs are detailed as follows.

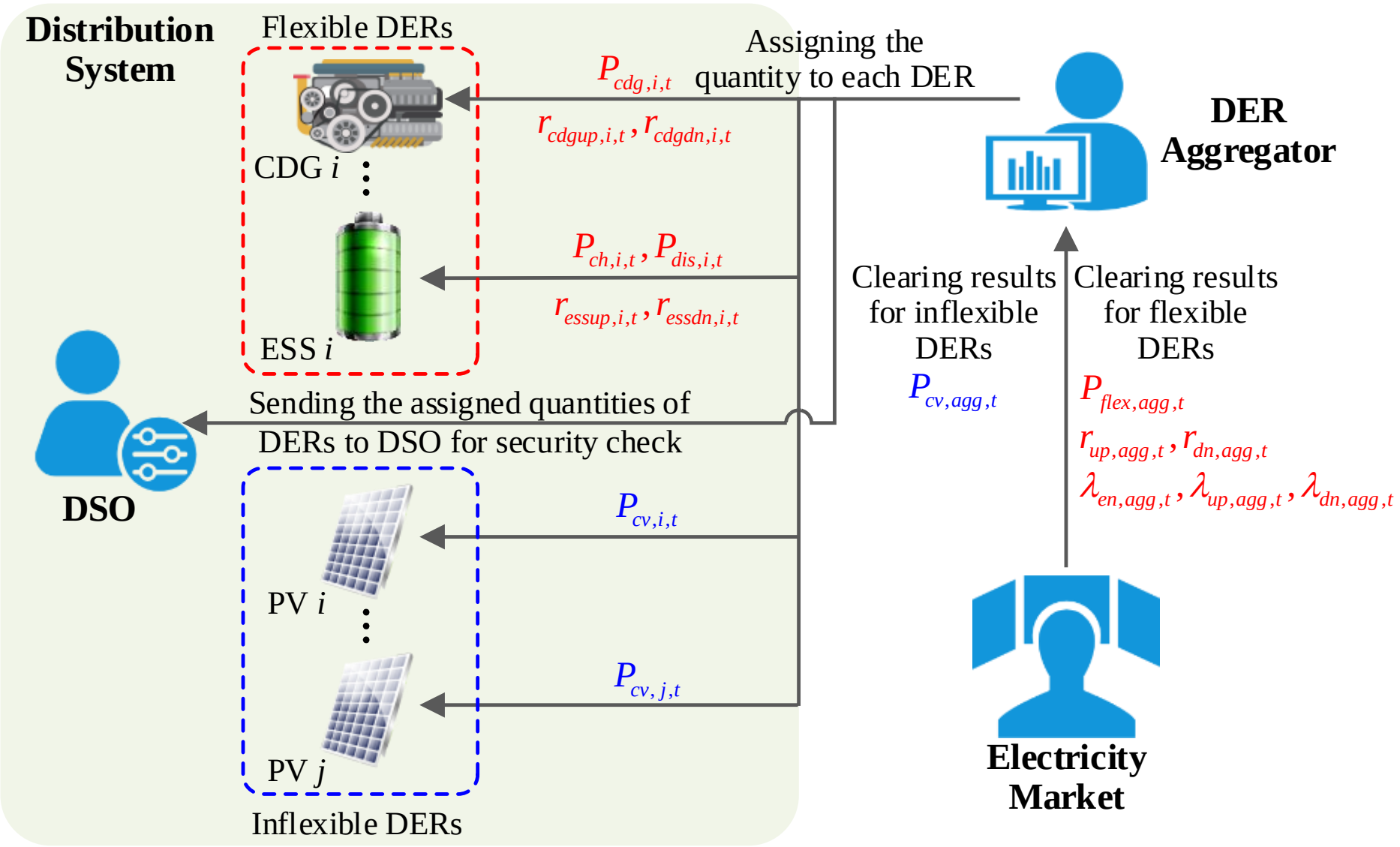

**Fig. 4.** Framework of the allocation of the scheduled quantities to each DER.

3) Allocation of the scheduled energy and reserve quantity constraints by the DER aggregator

$$P_{flex,agg,t} = \sum_{i\in N_{CDG}} P_{cdg,i,t} + \sum_{i\in N_{ESS}} \left(P_{dis,i,t} - P_{ch,i,t}\right), \tag{4a}$$

$$P_{cv,agg,t} = \sum_{i\in N_{PV}} P_{cv,i,t}^{NR} \,, \tag{4b}$$

$$r_{up,agg,t} = \sum_{i\in N_{CDG}} r_{cdgup,i,t} + \sum_{i\in N_{ESS}} r_{essup,i,t} \,, \tag{4c}$$

$$r_{dn,agg,t} = \sum_{i\in N_{CDG}} r_{cdgdn,i,t} + \sum_{i\in N_{ESS}} r_{essdn,i,t} \,. \tag{4d}$$

Eq. (4a) shows the allocation of the scheduled aggregated flexible energy quantity to the flexible DERs. Eq. (4b) indicates the allocation of the scheduled curtailment of aggregated inflexible energy quantity to PVs. Eqs. (4c) and (4d) describe the allocation of scheduled reserve quantities to the flexible DERs.

4) CDG constraints

$$0 \le P_{cdg,i,t} \le P_{cdg,i,\max}, \quad \forall i \in N_{CDG}, \tag{5a}$$

$$0 \le r_{cdgup,i,t} \le r_{cdgup,i,\max}, \quad \forall i \in N_{CDG}, \tag{5b}$$

$$0 \le r_{cdgdn,i,t} \le r_{cdgdn,i,\max}, \quad \forall i \in N_{CDG}, \tag{5c}$$

$$P_{cdg,i,t} + r_{cdgup,i,t} \le P_{cdg,i,\max}, \quad \forall i \in N_{CDG}, \tag{5d}$$

$$P_{cdg,i,\min} + r_{cdgdn,i,t} \le P_{cdg,i,t}, \quad \forall i \in N_{CDG}. \tag{5e}$$

Eqs. (5a)–(5c) limit the upper and lower bounds of the active power and reserve of CDG. Eqs. (5d) and (5e) enforce that the scheduled active power and reserve of CDG are within the output capacity.

5) ESS constraints

$$0 \le P_{ch,i,t} \le P_{ch,i,\max} I_{ch,i,t}, \quad \forall i \in N_{ESS}, \tag{6a}$$

$$0 \le P_{dis,i,t} \le P_{dis,i,\max} I_{dis,i,t}, \quad \forall i \in N_{ESS}, \tag{6b}$$

$$I_{ch,i,t} + I_{dis,i,t} \le 1, \quad \forall i \in N_{ESS}, \tag{6c}$$

$$E_{i,t} = E_{i,t-1} + \eta_{c,i} P_{ch,i,t} - \frac{1}{\eta_{d,i}} P_{dis,i,t}, \quad \forall i \in N_{ESS}, \tag{6d}$$

$$E_{i,\min} \le E_{i,t} \le E_{i,\max}, \quad \forall i \in N_{ESS}, \tag{6e}$$

$$E_{i,0} = E_{i,T}, \quad \forall i \in N_{ESS}, \tag{6f}$$

$$0 \le r_{essup,i,t} \le \min\left\{ P_{dis,i,\max} - P_{dis,i,t} + P_{ch,i,t}, \eta_{d,i}\left(E_{i,t-1} - E_{i,\min}\right)\right\}, \quad \forall i \in N_{ESS}, \tag{6g}$$

$$0 \le r_{essdn,i,t} \le \min\left\{ P_{ch,i,\max} + P_{dis,i,t} - P_{ch,i,t}, \frac{E_{i,\max} - E_{i,t-1}}{\eta_{c,i}} \right\}, \quad \forall i \in N_{ESS}. \tag{6h}$$

Eqs. (6a) and (6b) limit the upper and lower bounds of the charging and discharging power of ESS. Eq. (6c) ensures that charging and discharging do not proceed simultaneously. Eq. (6d) denotes the energy capacity of ESS by the end of time *t*. Eq. (6e) denotes the upper and lower limits of the energy capacity. Eq. (6f) guarantees that the energy capacities at the initial time and the final time remain identical. Eqs. (6g) and (6h) restrict the available upward and downward reserves of ESS.

6) PV constraints

$$0 \le P^{NR}_{cv,i,t} \le P_{av,i,t}, \quad \forall i \in N_{PV}. \tag{7a}$$

$$0 \le P^{UP}_{cv,i,t} \le \left(1 + \varepsilon_{PV}\right) P_{av,i,t}, \quad \forall i \in N_{PV}. \tag{7b}$$

$$0 \le P^{DN}_{cv,i,t} \le \left(1 - \varepsilon_{PV}\right) P_{av,i,t}, \quad \forall i \in N_{PV}. \tag{7c}$$

Eqs. (7a)–(7c) limit the active power curtailment of PV generation in normal scheduling scenarios, upward-reserve-activated scenarios, and downward-reserve-activated scenarios, respectively.

*2.2 Follower-level problem 1: Clearing process of day-ahead energy and reserve markets by ISO*

*2.2.1 Objective*

For the follower-level problem of the ISO, the objective is to maximize the social welfare (i.e., minimize the negative effects) of day-ahead energy and reserve markets, consisting of the cost of energy and reserve requirements for purchasing from the conventional generators and the DER aggregator.

$$\min \sum_{t\in T}\sum_{i\in N_G}\left(O_{g,i}P_{g,i,t} + O_{gup,i}r_{gup,i,t} + O_{gdn,i}r_{gdn,i,t}\right) + \sum_{t\in T}\left(o_{flex,agg,t}P_{flex,agg,t} + O_{cv,agg}P_{cv,agg,t} + o_{up,agg,t}r_{up,agg,t} + o_{dn,agg,t}r_{dn,agg,t}\right). \tag{8}$$

*2.2.2 Constraints*

$$P_{flex,agg,t} + P_{infl,agg,t} - P_{cv,agg,t} + \sum_{i\in N_G} P_{g,i,t} = \sum_{i\in N_D} P_{d,i,t} : \lambda_{en,t}, \tag{9a}$$

$$r_{up,agg,t} + \sum_{i\in N_G} r_{gup,i,t} = R_{up,t} + R_{up,agg,t} : \lambda_{up,t}, \tag{9b}$$

$$r_{dn,agg,t} + \sum_{i\in N_G} r_{gdn,i,t} = R_{dn,t} + R_{dn,agg,t} : \lambda_{dn,t}, \tag{9c}$$

$$r_{gdn,i,t} \le P_{g,i,t} \le P_{g,i,\max} - r_{gup,i,t} : \underline{v}_{g,i,t}, \overline{v}_{g,i,t}, \quad \forall i \in N_G, \tag{9d}$$

$$0 \le r_{gup,i,t} \le r_{gup,i,\max} : \; \underline{\nu}_{gup,i,t}, \bar{\nu}_{gup,i,t}, \quad \forall i \in N_G, \tag{9e}$$

$$0 \le r_{gdn,i,t} \le r_{gdn,i,\max} : \; \underline{\nu}_{gdn,i,t}, \bar{\nu}_{gdn,i,t}, \quad \forall i \in N_G, \tag{9f}$$

$$\underline{P}_{flex,agg,t} + r_{dn,agg,t} \le P_{flex,agg,t} \le \bar{P}_{flex,agg,t} - r_{up,agg,t} : \; \underline{\nu}_{flex,agg,t}, \bar{\nu}_{flex,agg,t}, \tag{9g}$$

$$0 \le r_{up,agg,t} \le \bar{r}_{up,agg,t} : \; \underline{\nu}_{up,agg,t}, \bar{\nu}_{up,agg,t}, \tag{9h}$$

$$0 \le r_{dn,agg,t} \le \bar{r}_{dn,agg,t} : \; \underline{\nu}_{dn,agg,t}, \bar{\nu}_{dn,agg,t}, \tag{9i}$$

$$0 \le P_{cv,agg,t} \le P_{infl,agg,t} : \; \underline{\nu}_{cv,agg,t}, \bar{\nu}_{cv,agg,t}, \tag{9j}$$

$$-P_{l,\max} \le G_{l,agg}\left(P_{flex,agg,t} + P_{infl,agg,t} - P_{cv,agg,t}\right) + \sum_{i \in N_G} G_{l,i} P_{g,i,t} - \sum_{i \in N_D} D_{l,i} P_{d,i,t} \le P_{l,\max} : \; \underline{\nu}_{l,t}, \bar{\nu}_{l,t}, \quad \forall l \in N_L. \tag{9k}$$

Eq. (9a) ensures the energy balance of the system, and its associated dual variable is the day-ahead energy price without transmission congestion. Eqs. (9b) and (9c) guarantee that the upward and downward reserve requirements are met, and their corresponding dual variables are the day-ahead clearing prices of upward and downward reserve. Eqs. (9d)–(9f) limit the available energy quantities and upward and downward reserve quantities of the conventional generators. Eqs. (9g)–(9i) restrict the available flexible energy quantities and upward and downward reserve quantities of the DER aggregator. Eq. (9j) ensures the curtailment of the aggregated inflexible supply. Eq. (9k) indicates the capacity limit of the transmission lines in the transmission system.

*2.3 Follower-level problem 2: Security check model by the DSO under different scenarios*

With the increasing penetration of DERs in the DS, the scheduling of DERs without considering the network constraints of the DS will significantly affect the secure operation of the DS, and even cause DS operation security problems, such as voltage problems and congestion problems. Thus, the scheduling scheme of each DER should be checked by DSO to guarantee the operation security of the DS. The SC problems of the DSO for the SC under three different scenarios, i.e., normal scheduling scenario, upward-reserve-activated scenario, and downward-reserve-activated scenario, are formulated to be the follower-level problems in this section.

*2.3.1 Objective*

The SC of the DS aims to minimize the distortion of voltage profiles:

$$\min \sum_{i \in N_{DS}} \left|V_{i,t}^s - 1\right|, \quad s \in \Omega. \tag{10}$$

*2.3.2 Constraints*

1) DER injected power equations under different scenarios

i) Normal scheduling scenario

$$P_{cdg,i,t}^{NR} = P_{cdg,i,t}, \quad \forall i \in N_{CDG}, \tag{11a}$$

$$P_{ess,i,t}^{NR} = P_{dis,i,t} - P_{ch,i,t}, \quad \forall i \in N_{ESS}, \tag{11b}$$

$$P_{pv,i,t}^{NR} = P_{av,i,t} - P_{cv,i,t}^{NR}, \quad \forall i \in N_{PV}, \tag{11c}$$

ii) Upward-reserve-activated scenario

$$P_{cdg,i,t}^{UP} = P_{cdg,i,t} + r_{cdgup,i,t}, \quad \forall i \in N_{CDG}, \tag{11d}$$

$$P_{ess,i,t}^{UP} = P_{dis,i,t} - P_{ch,i,t} + r_{essup,i,t}, \quad \forall i \in N_{ESS}, \tag{11e}$$

$$P_{pv,i,t}^{UP} = P_{av,i,t} + R_{dn,i,t} - P_{cv,i,t}^{UP}, \quad \forall i \in N_{PV}, \tag{11f}$$

iii) Downward-reserve-activated scenario

$$P_{cdg,i,t}^{DN} = P_{cdg,i,t} - r_{cdgdn,i,t}, \quad \forall i \in N_{CDG}, \tag{11g}$$

$$P_{ess,i,t}^{DN} = P_{dis,i,t} - P_{ch,i,t} - r_{essdn,i,t}, \quad \forall i \in N_{ESS}, \tag{11h}$$

$$P_{pv,i,t}^{DN} = P_{av,i,t} - R_{up,i,t} - P_{cv,i,t}^{DN}, \quad \forall i \in N_{PV}. \tag{11i}$$

Eqs. (11a)–(11c), (11d)–(11f), and (11g)–(11i) represent the scheduled energy quantity of CDG, ESS, and PV in the normal scheduling scenarios, upward-reserve-activated scenarios, and downward-reserve-activated scenarios, respectively.

2) Linear power flow equations of the DS

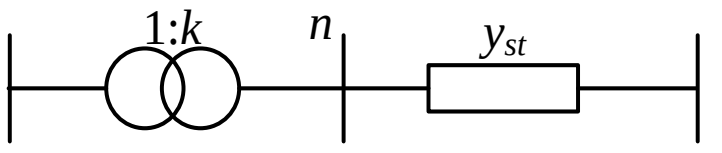

**Fig. 5.** Model of substation transformer branch with dummy bus.

As shown in Fig. 5, the substation transformer branch of the DS can be divided into an ideal transformer and a general branch by introducing a dummy bus $n$ [26]. According to [23], the nonlinear power flow equations of a DS can be approximately expressed as a set of linear equations. Based on the Big-M method [24], we extended the linearized power flow model in [23] to further consider the substation transformer tap, which is developed in Appendix A:

$$k_t^s V_{i,t}^s(t) \approx k_t^s e_{i,t}^s(t) = \left(k_t^s\right)^2 + \sum_{j \in N_{DS}} \left(R_{ij} P_{j,t}^s + X_{ij} Q_{j,t}^s\right), \ s \in \Omega, \tag{12a}$$

$$k_t^s \theta_{i,t}^s(t) \approx k_t^s f_{i,t}^s = \sum_{j \in N_{DS}} \left(X_{ij} P_{j,t}^s - R_{ij} Q_{j,t}^s\right), \ s \in \Omega, \tag{12b}$$

$$P_{i,t}^s = P_{cdg,i,t}^s + P_{pv,i,t}^s + P_{ess,i,t}^s - P_{d,i,t}, \ s \in \Omega, \tag{12c}$$

$$Q_{i,t}^s = Q_{dg,i,t}^s + Q_{pv,i,t}^s + Q_{cb,i,t}^s - Q_{d,i,t}, \ s \in \Omega. \tag{12d}$$

Eqs. (12a) and (12b) represent power flow equations of the DS. Eqs. (12c) and (12d) respectively denote the net injected active and reactive power of buses in the DS.

3) Limits on reactive power output to CDGs in the DS

$$-P_{cdg,i,t}^s \tan\varphi_{cdg,i} \le Q_{cdg,i,t}^s \le P_{cdg,i,t}^s \tan\varphi_{cdg,i}, \ \forall i \in N_{CDG}, \ s \in \Omega. \tag{13}$$

4) Limits on reactive power output to PVs in the DS

$$-P_{pv,i,t}^s \tan\varphi_{pv,i} \le Q_{pv,i,t}^s \le P_{pv,i,t}^s \tan\varphi_{pv,i}, \ \forall i \in N_{PV}, \ s \in \Omega. \tag{14}$$

5) Operation constraints of CBs in the DS

$$Q_{cb,i,t}^s = Q_{cb,i,\min}^s + \Delta Q_{cb,i}^s C_{cb,i,t}^s, \ \forall i \in N_{CB}, \ s \in \Omega, \tag{15a}$$

$$Q_{cb,i,\min}^s \le Q_{cb,i,t}^s \le Q_{cb,i,\max}^s, \ \forall i \in N_{CB}, \ s \in \Omega. \tag{15b}$$

6) Operation constraints of substation transformer in the DS

$$k_t^s = k_{\min} + \Delta k \sum_{n \in N_k} 2^n w_{n,t}^s, \ w_{n,t}^s \in \{0,1\}, \ s \in \Omega, \tag{16a}$$

$$k_{\min} \le k_t^s \le k_{\max}, \ s \in \Omega. \tag{16b}$$

7) Limits on bus voltage in the DS

$$V_{i,\min} \le V_{i,t}^s \le V_{i,\max}, \ \forall i \in N_{DS}, \ s \in \Omega. \tag{17}$$

8) Limits on line current in the DS

$$\left|I_{ij,t}^s\right|^2 = \left(I_{re,ij,t}^s\right)^2 + \left(I_{im,ij,t}^s\right)^2 = \left|y_{ij}\right|^2 \left|V_{i,t}^s e^{j\theta_{i,t}^s} - V_{j,t}^s e^{j\theta_{j,t}^s}\right|^2 \le I_{ij,\max}^2, \ \forall ij \in N_{DL}, \ s \in \Omega. \tag{18}$$

To decrease the computational complexity, Eq. (18) could be approximated using the following outer-approximation constraints, similar to [27]:

$$-I_{ij,\max} \le I_{re,ij,t}^s \le I_{ij,\max}, \ \forall ij \in N_{DL}, \ s \in \Omega, \tag{19a}$$

$$-I_{ij,\max} \le I_{im,ij,t}^s \le I_{ij,\max}, \ \forall ij \in N_{DL}, \ s \in \Omega, \tag{19b}$$

$$-\sqrt{2} I_{ij,\max} \le I_{re,ij,t}^s + I_{im,ij,t}^s \le \sqrt{2} I_{ij,\max}, \ \forall ij \in N_{DL}, \ s \in \Omega, \tag{19c}$$

$$-\sqrt{2} I_{ij,\max} \le I_{re,ij,t}^s - I_{im,ij,t}^s \le \sqrt{2} I_{ij,\max}, \ \forall ij \in N_{DL}, \ s \in \Omega, \tag{19d}$$

where the real and imaginary parts of the line current can be expressed as the following linearized form:

$$I_{re,ij,t}^s = g_{ij}\left(V_{i,t}^s - V_{j,t}^s\right) - b_{ij}\left(\theta_{i,t}^s - \theta_{j,t}^s\right), \ \forall ij \in N_{DL}, \ s \in \Omega, \tag{20a}$$

$$I_{im,ij,t}^s = b_{ij}\left(V_{i,t}^s - V_{j,t}^s\right) + g_{ij}\left(\theta_{i,t}^s - \theta_{j,t}^s\right), \ \forall ij \in N_{DL}, \ s \in \Omega. \tag{20b}$$

## 3 Solution and methodology

Mathematically, the proposed Stackelberg game model illustrated above is essentially a bi-level mixed-integer nonlinear programming model with both linear programming (LP) and MILP follower-level problems, which cannot be solved directly using any off-the-shelf optimization solver. In this section, we detail the solution of the proposed Stackelberg game model.

### *3.1 Transformation based on KKT-based reformulation approach*

In the proposed Stackelberg game model, the follower-level problem of the ISO is an LP model. Thus, the follower-level problem of the ISO can be equivalently substituted with its corresponding KKT condition [28]. After replacement, the bilinear terms in the leader-level objective function and complementary constraints of the KKT conditions can be easily linearized based on the strong duality theorem [28] and the Big-M method [24, 29]. Consequently, we can obtain an equivalent bi-level mixed-integer linear programming (BMILP) model with only MILP follower-level problems. The transformation process of the proposed Stackelberg game model is illustrated in Fig. 6.

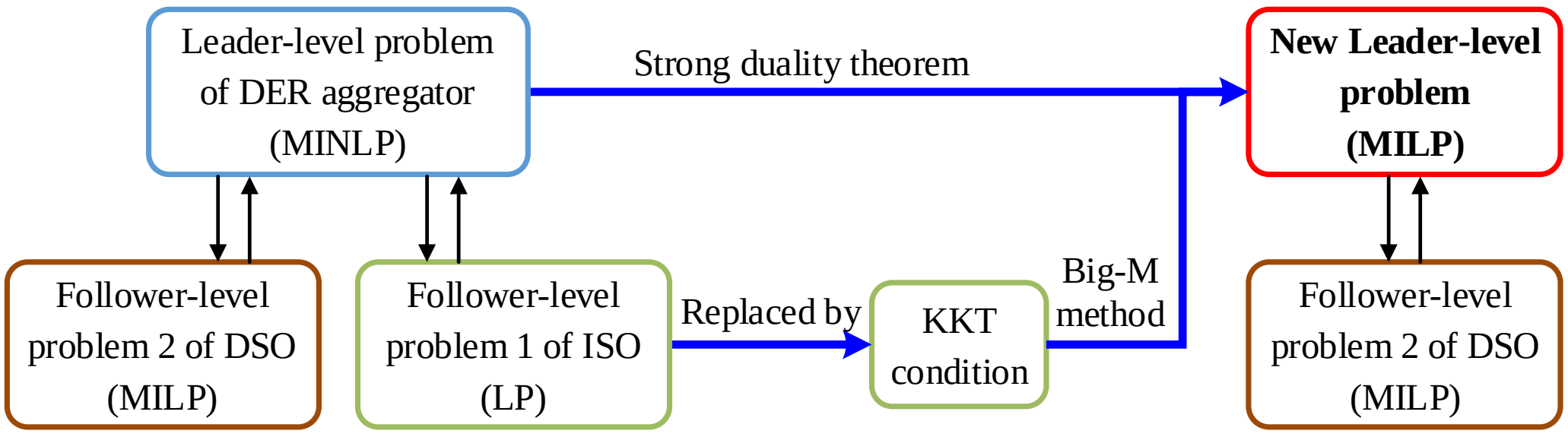

**Fig. 6.** Transformation of the proposed Stackelberg game model with KKT-based reformulation approach.

*3.2 RBRD algorithm*

After replacement and linearization, the proposed Stackelberg game model is transformed into an equivalent BMILP model with only MILP follower-level problems. To simplify the discussion of the solution, we rewrite the BMILP model into the following compact form:

$$\min \quad F_{UL} = \boldsymbol{c}_{UL}^T \boldsymbol{x} + \sum_{s\in\Omega}\sum_{t\in T}\left(\boldsymbol{d}_{UL}^s\right)^T \boldsymbol{m}_t^s \,, \tag{21a}$$

$$\text{s.t.} \quad \boldsymbol{A}_{UL}\boldsymbol{x} \le \boldsymbol{b}_{UL} \,, \tag{21b}$$

$$\boldsymbol{m}_t^s \in \arg\min_{\boldsymbol{m}_t^s,\boldsymbol{z}_t^s}\left\{ f_{LL,t}^s = \left(\boldsymbol{d}_{LL,t}^s\right)^T \boldsymbol{m}_t^s : \boldsymbol{A}_{LL,t}^s\boldsymbol{x} + \boldsymbol{B}_{LL,t}^s\boldsymbol{m}_t^s + \boldsymbol{C}_{LL,t}^s\boldsymbol{z}_t^s \le \boldsymbol{b}_{LL,t}^s \right\}, \quad \forall t\in T, \quad \forall s\in\Omega \,, \tag{21c}$$

where Eq. (21b) represents the constraints of the new MILP leader-level problem, and Eq. (21c) represents the MILP follower-level problem for the SC by the DSO.

Due to the nonconvexity, the MILP follower-level problems (21c) cannot be directly replaced with the corresponding KKT conditions [30]. By extending the method proposed in [31], an RBRD algorithm [25] is proposed to globally solve the BMILP problem with multiple MILP follower-level problems. Therefore, we apply the RBRD algorithm to solve the BMILP model (21). Following [25], we can formulate the master problem and subproblems as follows.

Master problem (denoted as **MP**):

$$\min \quad F_{UL} = \boldsymbol{c}_{UL}^T \boldsymbol{x} + \sum_{s\in\Omega}\sum_{t\in T}\left(\boldsymbol{d}_{UL}^s\right)^T \boldsymbol{m}_t^s \,, \tag{22a}$$

$$\text{s.t.} \quad \boldsymbol{A}_{UL}\boldsymbol{x} \le \boldsymbol{b}_{UL} \,, \tag{22b}$$

$$\boldsymbol{A}_{LL,t}^s\boldsymbol{x} + \boldsymbol{B}_{LL,t}^s\boldsymbol{m}_t^s + \boldsymbol{C}_{LL,t}^s\boldsymbol{z}_t^s \le \boldsymbol{b}_{LL,t}^s, \quad \forall t\in T, \quad \forall s\in\Omega \,, \tag{22c}$$

$$\begin{aligned} \left(\boldsymbol{d}_{LL,t}^s\right)^T \boldsymbol{m}_t^s \le \min_{\boldsymbol{m}_{t,c,l}^s,\boldsymbol{\tau}_{t,c,l}^s} & \left\{ f_{LL,t}^s = \left(\boldsymbol{d}_{LL,t}^s\right)^T \boldsymbol{m}_{t,c,l}^s + \left(\boldsymbol{\rho}_t^s\right)^T \boldsymbol{\tau}_{t,c,l}^s : \right. \\ & \left. \boldsymbol{A}_{LL,t}^s\boldsymbol{x} + \boldsymbol{B}_{LL,t}^s\boldsymbol{m}_{t,c,l}^s + \boldsymbol{C}_{LL,t}^s\boldsymbol{z}_{t,c,l}^s - \boldsymbol{D}_t^s\boldsymbol{\tau}_{t,c,l}^s \le \boldsymbol{b}_{LL,t}^s, \boldsymbol{\tau}_{t,c,l}^s \ge \boldsymbol{0} \right\}, \quad \forall t\in T, \quad \forall s\in\Omega, \quad \forall l\in\left\{1,\ldots,p_t^s\right\} \end{aligned} . \tag{22d}$$

Subproblem at time $t$ in scenario $s$ at given leader-level decision $\boldsymbol{x}^*$ (denoted as $\mathbf{SP}_t^s$ ):

$$\min_{\boldsymbol{m}_t^s,\boldsymbol{z}_t^s} \quad f_{LL,t}^s\left(\boldsymbol{x}^*\right) = \left(\boldsymbol{d}_{LL,t}^s\right)^T \boldsymbol{m}_t^s, \quad t\in T, \quad s\in\Omega \,, \tag{23a}$$

$$\text{s.t.} \quad \boldsymbol{A}_{LL,t}^s\boldsymbol{x}^* + \boldsymbol{B}_{LL,t}^s\boldsymbol{m}_t^s + \boldsymbol{C}_{LL,t}^s\boldsymbol{z}_t^s \le \boldsymbol{b}_{LL,t}^s, \quad t\in T, \quad s\in\Omega \,. \tag{23b}$$

*3.3 Accelerated method*

The master problem (22) is a BMILP model with only LP follower-level problems, and it can be solved using the KKT-based reformulation approach [28]. However, new LP follower-level problems (22d) will be added to the master problem at each iteration. Consequently, a large number of binary variables should be introduce to linearize the complementary constraints of the corresponding KKT conditions of the LP follower-level problems, which will significantly increase the complexity of the master problem.

The numbers of binary variables introduced to linearize the complementary constraints are equal to the numbers of inequality constraints in (22d). Considering the characteristic of problems (22d), most inequlity constraints (e.g., constraints of bus voltage and line current) are inactive. Thus, identifying and removing these inactive inequality constraints before adding new LP follower-level problems to the master problem will significantly reduce the numbers of binary variables and accelerate the solution.

For each inequality constraint in (22d), the upper bound $b_{LL,t,j}^{s,u}$ is determined via the following bi-level problem:

$$b_{LL,t,j}^{s,u} = \max \quad \boldsymbol{A}_{LL,t,j}^s\boldsymbol{x} + \boldsymbol{B}_{LL,t,j}^s\boldsymbol{m}_{t,c,l}^s + \boldsymbol{C}_{LL,t,j}^s\boldsymbol{z}_{t,c,l}^s - \boldsymbol{D}_t^s\boldsymbol{\tau}_{t,c,l}^s, \quad t\in T, \quad s\in\Omega, \quad l\in\left\{1,\ldots,p_t^s\right\}, \tag{24a}$$

s.t. $A_{UL}x \le b_{UL}$ , (24b)

$$\left(m_t^s, \tau_{t,c,l}^s\right) \in \min_{m_{t,c,l}^s, \tau_{t,c,l}^s} \left\{ f_{LL,t}^s = \left(d_{LL,t}^s\right)^T m_{t,c,l}^s + \left(\rho_t^s\right)^T \tau_{t,c,l}^s : A_{LL,t}^s x + B_{LL,t}^s m_{t,c,l}^s + C_{LL,t}^s z_{t,c,l}^s - D_t^s \tau_{t,c,l}^s \le b_{LL,t}^s, \tau_{t,c,l}^s \ge \mathbf{0} \right\},\ t \in T,\ s \in \Omega,\ l \in \left\{1, \dots, p_t^s\right\}, \tag{24c}$$

where the subscript $j$ denotes the $j$th inequality constraint. Similarly, we can obtain the upper and lower bounds of slack variables $\tau_{t,c,l}^s$ .

Then, we can judge that the inequality constraints are inactive if $b_{LL,t,j}^s > b_{LL,t,j}^{s,u}$ . In addition, the slack variables $\tau_{t,c,l}^s$ can be removed if both its upper and lower bounds are zero. Obviously, problem (24) for the follower-level inequality constraints are independent of each other and thus can be solved simultaneously in parallel. It should be noted that problem (24) can be further simplified by only involving the leader-level constraints of the couple variables between leader- and follower-level model (22d). The calculation flowchart of the A-RBRD algorithm is depicted in Fig. 7.

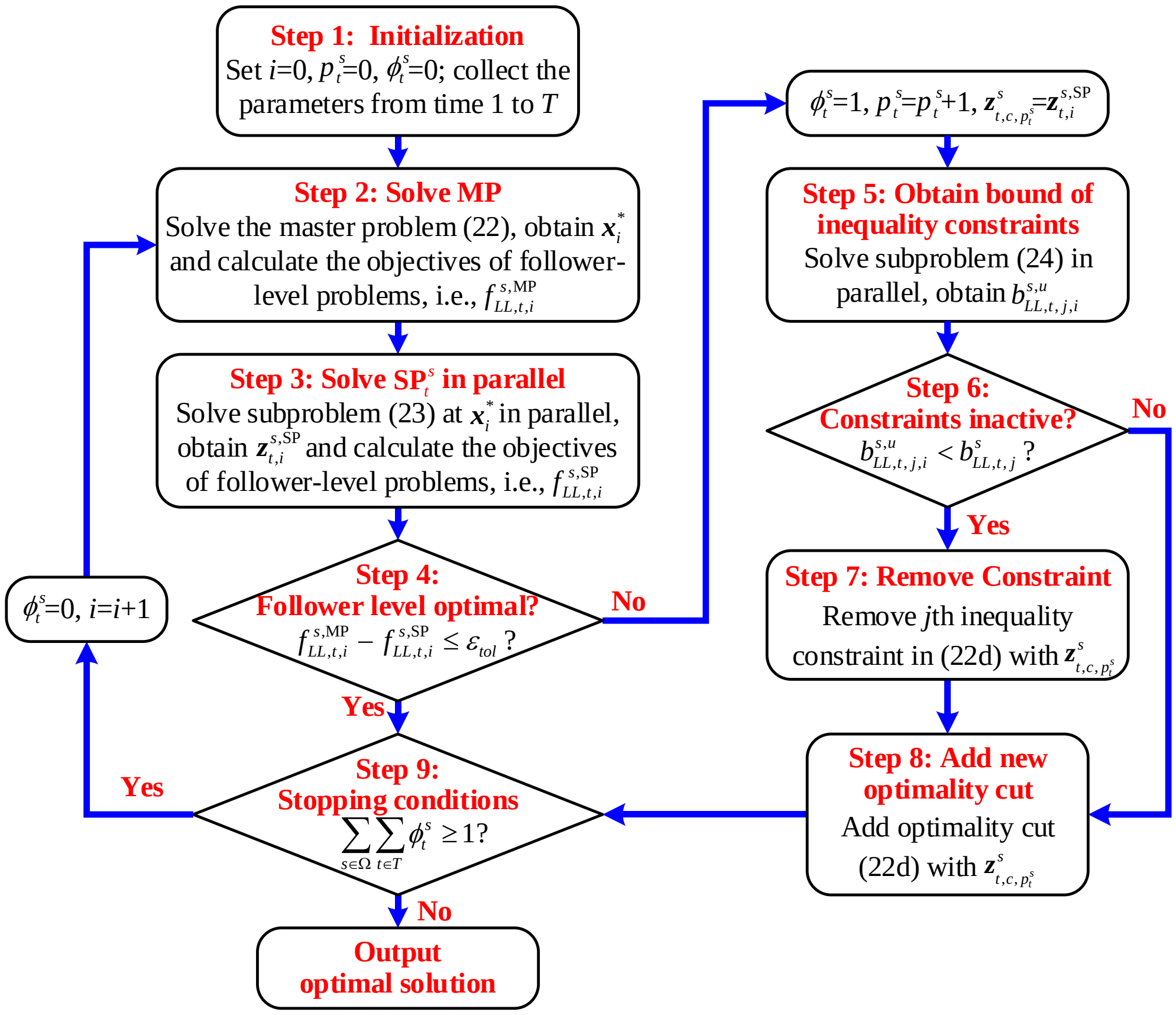

**Fig. 7.** Flowchart of the proposed A-RBRD algorithm for the proposed Stackelberg game model.

## 4 Case studies and analysis

In this section, the effectiveness of the proposed Stackelberg game model for the DER aggregator in day-ahead energy and reserve market considering SC was verified on a constructed integrated T&D system and a practical system in China. The proposed Stackelberg game model was programmed using MATLAB and solved using GUROBI. The simulation was done on a computer equipped with an Intel Xeon E3-1245 v3 CPU and 16 GB of RAM.

*4.1 Modified IEEE 30-bus transmission system with 18-bus DS*

First, we built an integrated T&D system based on a modified IEEE 30-bus transmission system [32] and a modified 18-bus DS [33] for simulation, as shown in Fig. 8. There were two CDGs, two PVs, two ESSs, and one CB in the DS. The basis capacity was set to 1 MVA. The adjustment range of the substation transformer tap ratio in the DS was 1 ± 8 × 1.25%. The CB had five groups and a reactive power capacity of 1 Mvar. The maximum active power output and maximum upward and downward reserve capacities of CDG were 3 MW, and the power factor was 0.9. The ESS possessed a maximum charging/discharging power of 400 kW with a capacity of 1200 kWh. Both the charging and discharging efficiency of ESS were 0.95. The power factor of PV was 0.85, and the prediction error of PV active power was $\varepsilon_{PV}$=0.2. The cost parameters of DERs were as follows: $b_{cdgp,i} = 150$ ¥/MWh, $b_{cdgq,i}^{NR} = 15$ ¥/Mvarh, $b_{cdgq,i}^{UP} = b_{cdgq,i}^{DN} = 1.875$ ¥/Mvarh, $b_{cv,i}^{NR} = 100$ ¥/MWh, $b_{cv,i}^{UP} = b_{cv,i}^{DN} = 12.5$ ¥/MWh, $b_{pvq,i}^{NR} = 10$ ¥/MWh, $b_{pvq,i}^{UP} = b_{pvq,i}^{DN} = 1.25$ ¥/MWh, $b_{ch,i} = 3$ ¥/MWh, and $b_{dis,i} = 3$ ¥/MWh. The loads of the DS and the available active power output of PVs are illustrated in Fig. 9.

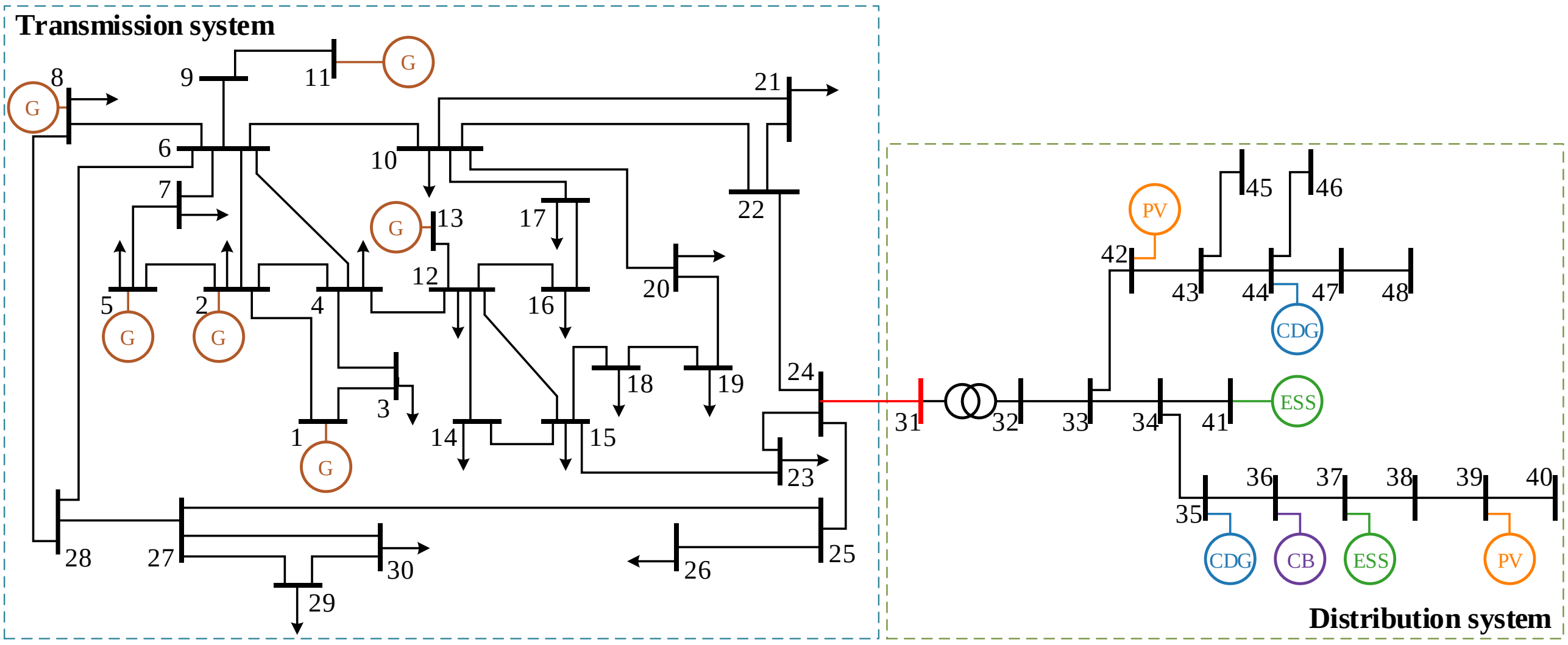

**Fig. 8.** Schematic of the constructed integrated T&D system.

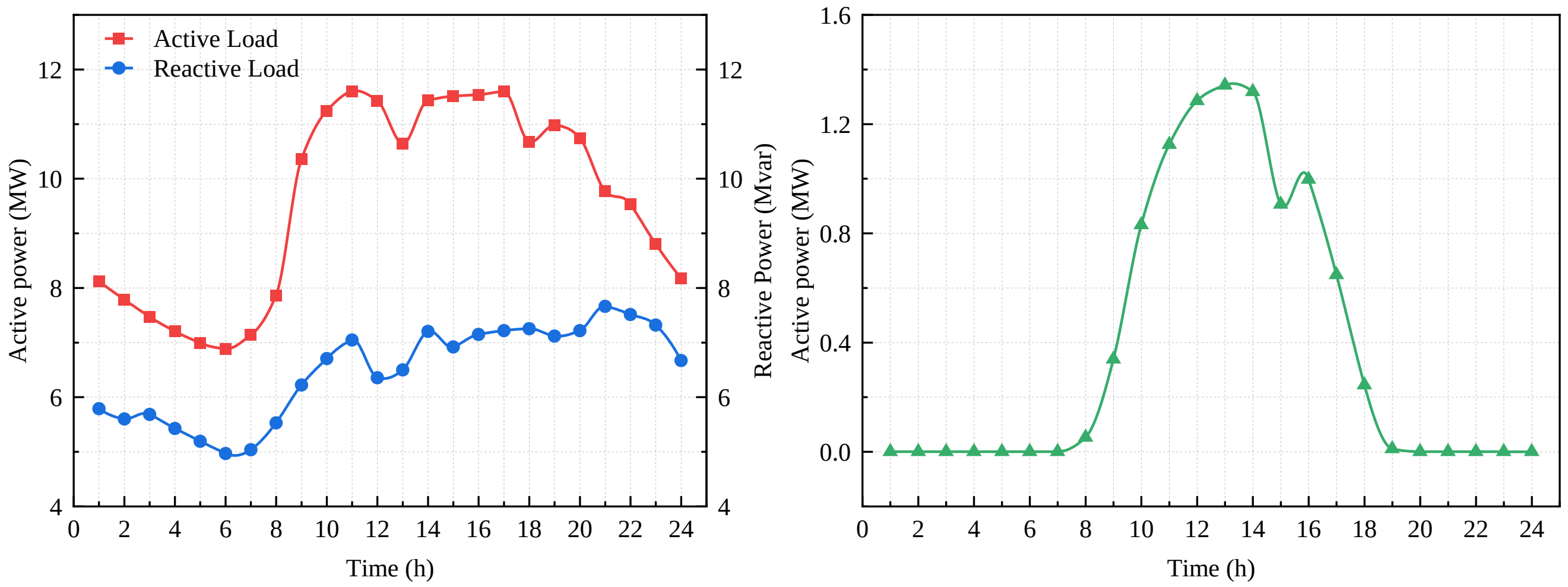

(a) Daily Loads of DS (b) Available active power output of PV

**Fig. 9.** Loads and available active power output of PVs in the DS.

To verify the effectiveness of the proposed Stackelberg game model for the DER aggregator participating in day-ahead energy and reserve markets considering the SC by the DSO, we considered the following three cases:

- Case 1: The DER aggregator participates in day-ahead energy and reserve markets without the SC.
- Case 2: The DER aggregator participates in day-ahead energy and reserve markets considering the SC, the upper and lower voltage magnitude limits are set to 0.96 p.u. and 1.04 p.u., and congestion is not considered.
- Case 3: The DER aggregator participates in day-ahead energy and reserve markets considering the SC, the upper and lower bus voltage magnitude limits are set to 0.96 p.u. and 1.04 p.u., and the current capacity of line 33–34 is set to 6.0 p.u.

It should be noted that Case 1 is the traditional Stackelberg game between DER aggregator and electricity market without considering the SC of DERs by the DSO, which serves as a benchmark.

Table 1 presents the total profits and the breakdown of these profits under Cases 1–3 in the constructed integrated T&D system. Referring to Table 1, the total profits of the DER aggregator in Cases 2 and 3 are smaller than that in Case 1. This demonstrates that the profits and available energy and reserve services of the DER aggregator in day-ahead market are limited by the secure operation constraints of the DS. Further observing the breakdown of these profits, we find that the profits of DERs in the energy market and upward reserve market are similar. The biggest differences in profits among these three cases occur in the downward reserve market. With the voltage magnitude limits and line current limits in the DS, the profits of the DER aggregator in the downward reserve market decrease, especially the profits of CDGs. The profits of ESSs in the downward reserve market in Cases 1 and 2 are similar, while those in Case 3 are slightly lower. The upward reserve capacity offered by the DER aggregator is provided by ESSs in the DS.

**Table 1:** Profits of the DER aggregator under three cases in the constructed integrated T&D system

| | Case 1 | Case 2 | Case 3 |
|---|---|---|---|
| Total profits (¥) | 39284.952 | 35981.592 | 35180.472 |
| Profits in energy market (¥) | 23859.134 | 23853.378 | 23833.379 |
| Profits in upward reserve market (¥) | 2131.065 | 2132.131 | 2140.162 |
| Profits in downward reserve market (¥) | 13294.752 | 11314.349 | 10533.360 |
| Costs of reactive power output (¥) | / | 1318.266 | 1326.430 |
| Profits of CDGs in energy market (¥) | 18643.154 | 18643.154 | 18643.154 |
| Profits of CDGs in upward reserve market (¥) | 0 | 0 | 0 |
| Profits of CDGs in downward reserve market (¥) | 11728.765 | 9791.448 | 9082.163 |
| Costs of reactive power output of CDGs (¥) | / | 1177.266 | 1185.430 |
| Profits of ESSs in energy market (¥) | -7.376 | -13.132 | -33.131 |
| Profits of ESSs in upward reserve market (¥) | 2131.065 | 2132.131 | 2140.241 |
| Profits of ESSs in downward reserve market (¥) | 1565.987 | 1522.901 | 1451.198 |
| Profits of PVs in energy market (¥) | 5223.356 | 5223.356 | 5223.356 |
| Costs of reactive power output of PVs (¥) | / | 141.000 | 141.000 |

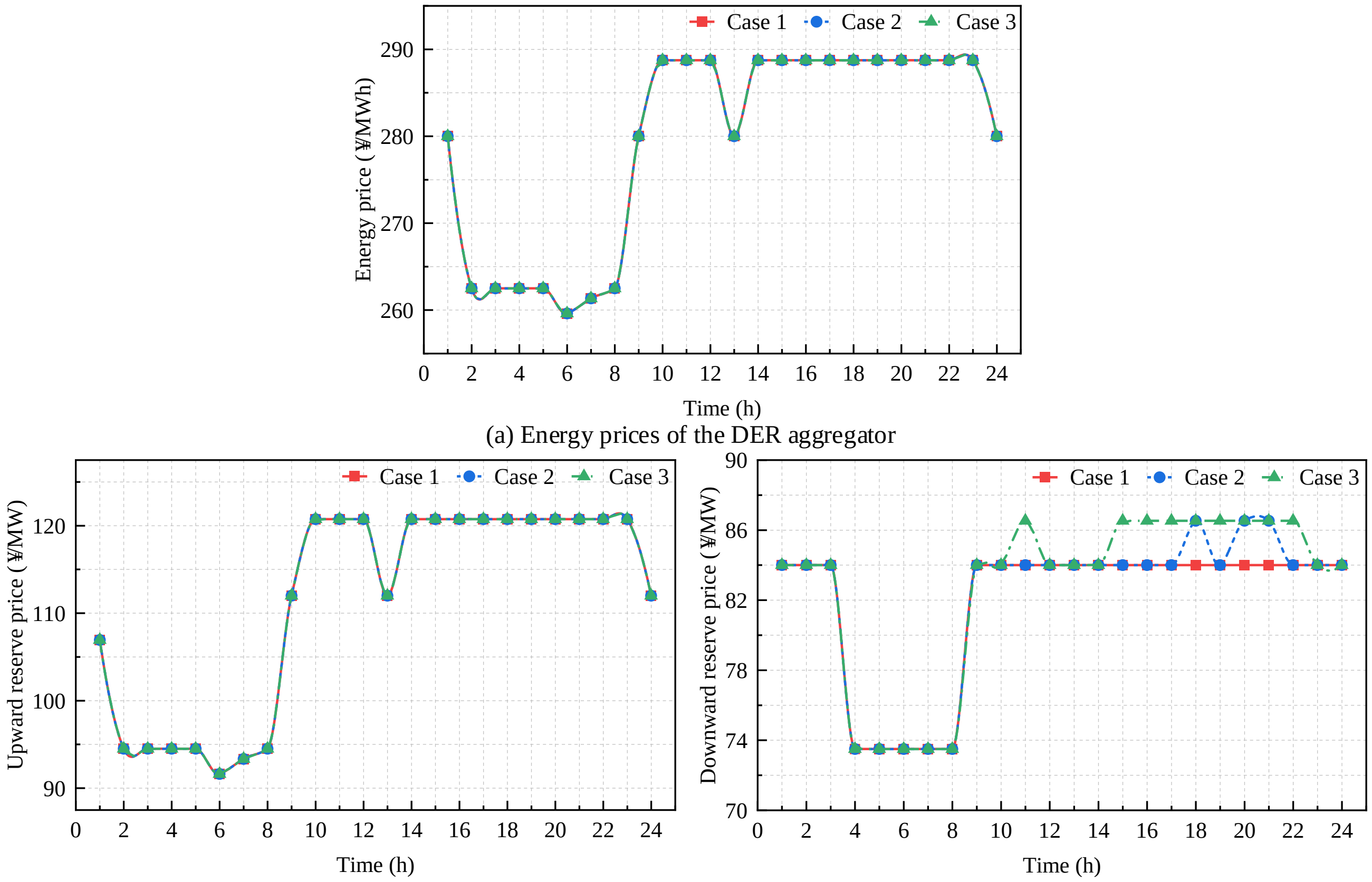

(a) Energy prices of the DER aggregator

(b) Upward reserve prices of the DER aggregator

(c) Downward reserve prices of the DER aggregator

**Fig. 10.** Comparison of the clearing prices of downward reserve market under Cases 1–3.

Fig. 10(a)–(c) compare the energy, upward reserve, and downward reserve clearing prices of day-ahead energy and reserve markets under Cases 1–3. We can observe that the energy and upward reserve prices under Cases 1–3 are identical, whereas the downward reserve prices become higher under Cases 1–3.

To further analyze the difference in the downward reserve market under Cases 1–3, the scheduled downward reserve of the DER aggregator and their allocation to CDGs and ESSs are illustrated in Fig. 11. Compared to Case 1, the scheduled downward reserve capacities of the DER aggregator become smaller under Cases 2 and 3. The reason is that the available downward reserve capacities of flexible DERs are limited by the voltage magnitude and line current limits in the DS. Thus, the integrated T&D system is forced to purchase downward reserve with higher costs, leading to higher downward reserve prices, as shown in Fig. 10(c). We observe from Fig. 11(c) that the downward reserve capacities provided by ESSs change significantly in Case 3, when the current limits of lines 33–34 are involved.

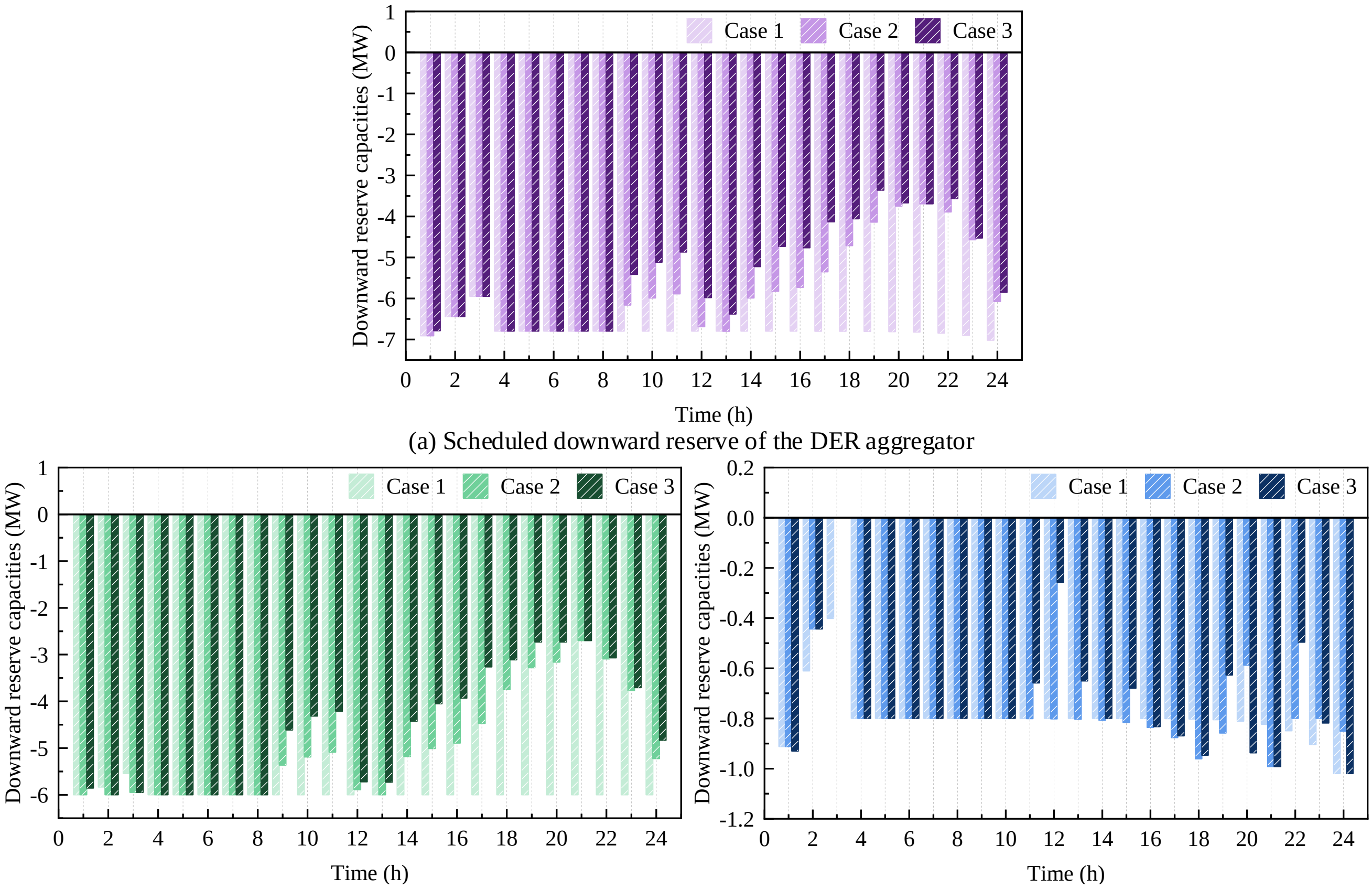

(a) Scheduled downward reserve of the DER aggregator

(b) Scheduled downward reserve of all DGs

(c) Scheduled downward reserve of all ESSs

**Fig. 11.** Comparison of the scheduled downward reserve of the DER aggregator and the allocation to CDGs and ESSs under Cases 1–3.

### *4.2 Practical 431-bus transmission system with 42-bus distribution system*

To further analyze the effectiveness of the proposed Stackelberg game model in practical applications, we conducted a simulation on a practical system consisting of a 431-bus transmission system with 45 conventional generators and a 42-bus DS. There were four CDGs, four PVs, four ESSs, and two CBs in the DS. Each CB had five groups and a reactive power capacity of 0.5 Mvar. The ESS possessed a maximum charging/discharging power of 500 kW with a capacity of 1500 kWh. The other parameters were identical to the constructed system in Section 4.1. The loads of the DS and the available active power of PVs are illustrated in Fig. 12. In this section, we studied the following two cases to verify the proposed Stackelberg game model:

- Case 4: The DER aggregator participates in day-ahead energy and reserve markets without the SC.
- Case 5: The DER aggregator participates in day-ahead energy and reserve markets considering the SC.

**Table 2:** Profits of the DER aggregator under two cases in the practical system

| | **Case 4** | **Case 5** |
|---|---|---|
| Total profits (¥) | 201091.18 | 194568.52 |
| Profits in energy market (¥) | 159813.28 | 160334.95 |
| Profits in upward reserve market (¥) | 9035.86 | 8862.94 |
| Profits in downward reserve market (¥) | 32242.04 | 27510.67 |
| Costs of reactive power output (¥) | / | 2140.04 |

Table 2 details the total profits of the DER aggregator and the breakdown of these profits under Cases 4 and 5 in the practical system. As shown in Table 2, the total profits in Case 4 are higher than those in Case 5. Similar to the results of the constructed system, the main difference in profits among Cases 4 and 5 occurs in the downward reserve market. Fig. 13 depicts the comparison of the clearing results in the reserve market between Cases 4 and 5. We observe that the downward reserve prices in Case 5 at time periods 10 and 18 are higher than in Case 4, whereas the upward reserve prices are identical. The available downward reserve capacities offered in Case 5 are smaller due to the network constraints in the DS. In addition, the scheduled upward reserve capacities of DERs are quite different between Cases 4 and 5, although the profits are similar.

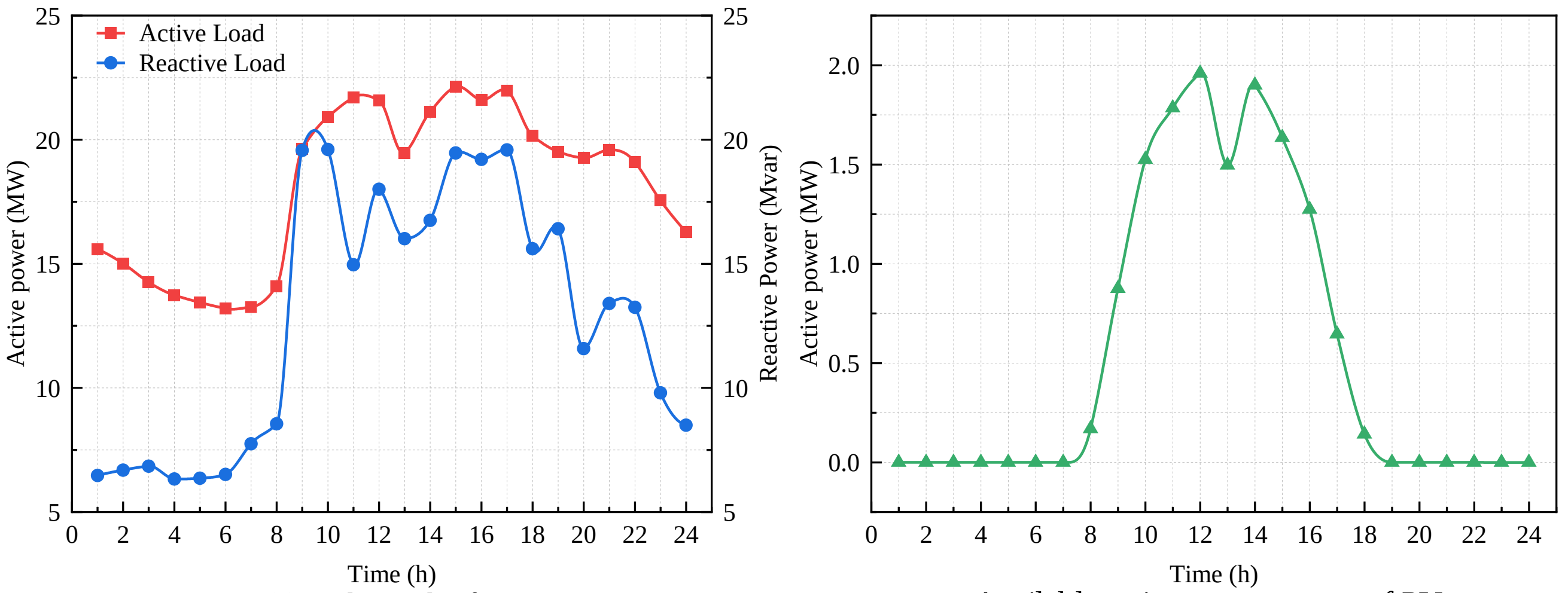

(a) Daily Loads of DS (b) Available active power output of PV

**Fig. 12.** Loads and available active power output of PVs in practical 42-bus DS.

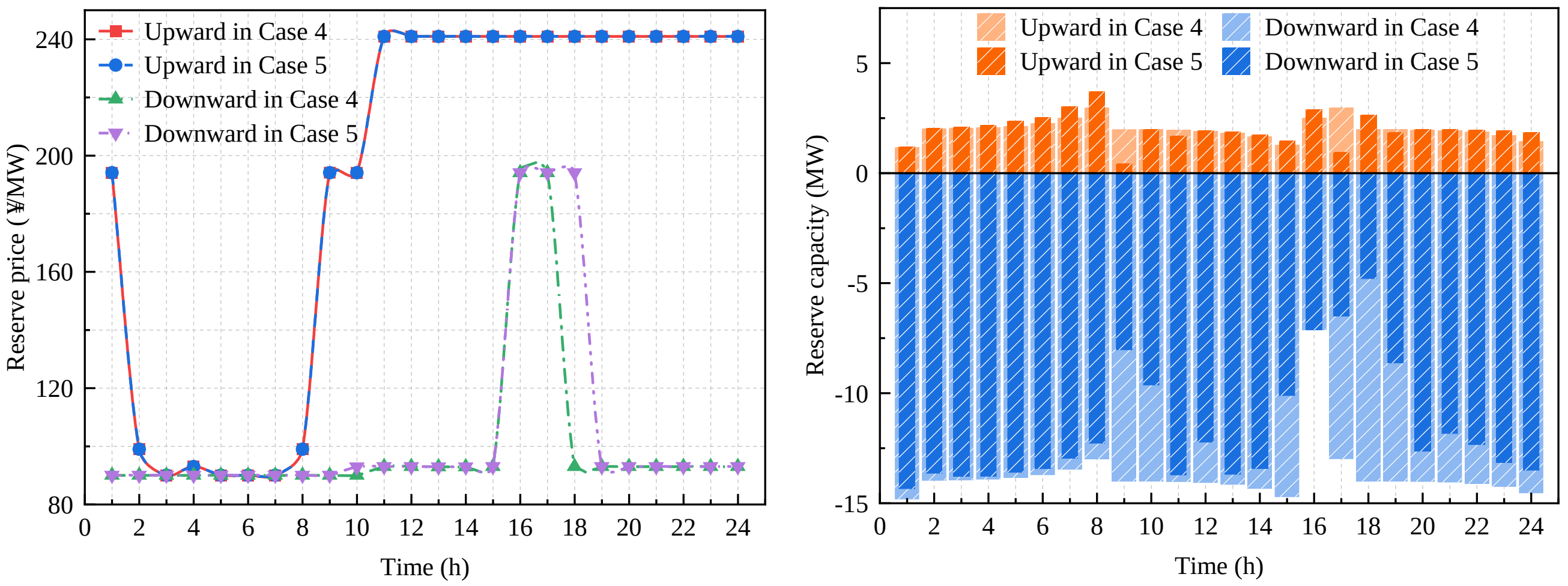

(a) Comparisons of reserve prices under Cases 4 and 5 (b) Comparisons of scheduled reserve under Cases 4 and 5

**Fig. 13.** Comparison of the clearing results of the reserve market under Cases 4 and 5.

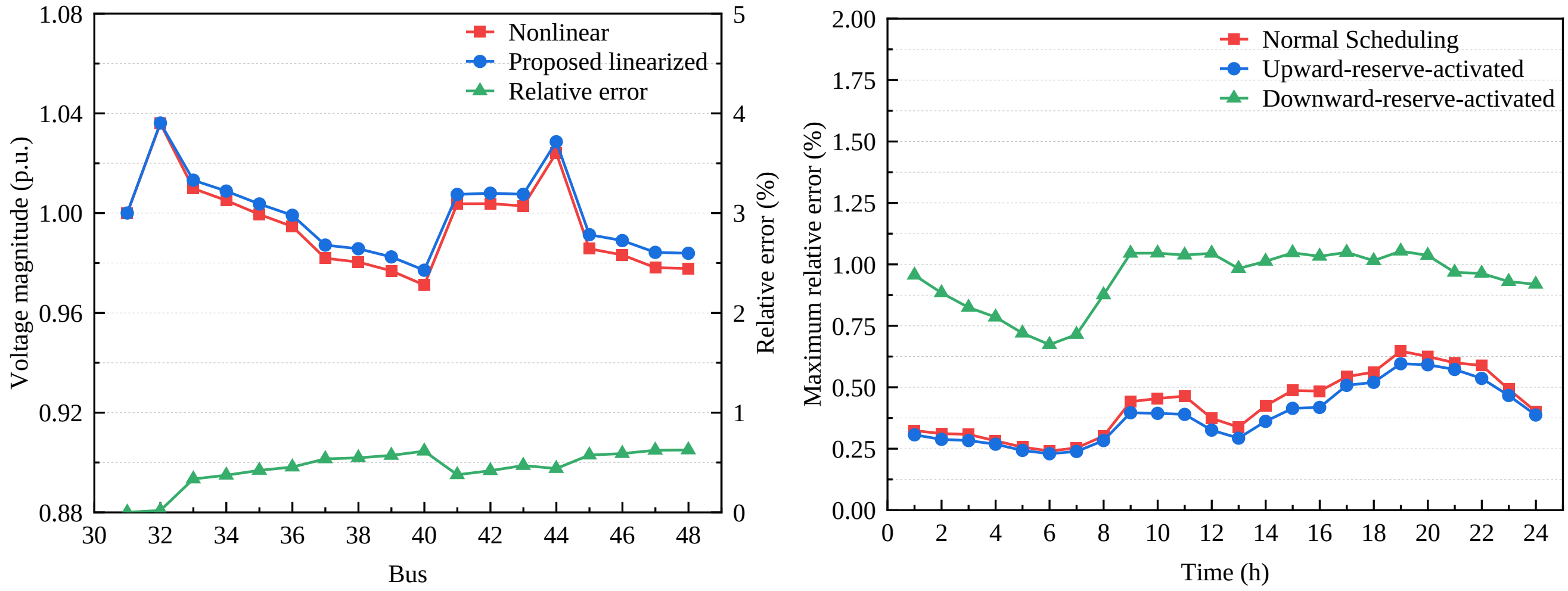

(a) Voltage magnitudes of DS at time period 20 under normal scheduling scenario obtained by two power flow models and the corresponding relative errors

(b) Maximum relative errors of the voltage magnitudes in DS at all time periods under three different scenarios

**Fig. 14.** Voltage magnitudes and relative errors in the DS obtained by the nonlinear and the proposed linearized power flow models.

*4.3 Accuracy of the proposed linearized power flow model for the DS considering substation transformer tap*

In this subsection, we take the simulation results of Case 3 as an example to analyze the accuracy of the proposed linearized power flow model for the DS considering substation transformer tap. The original nonlinear power flow model serves as a benchmark. Fig. 14(a) depicts the bus voltage magnitudes in the DS obtained by the nonlinear and the proposed linearized power flow model at time period 20 in the normal scheduling scenario, and the corresponding relative errors are also given. Fig. 14(b) presents the maximum relative errors of the bus voltage magnitudes in DS under three different scenarios. We can observe that the relative errors of the proposed linearized power flow model are smaller than 1.1%. The relative errors of those buses are larger when their electrical distances are further from the root bus of the DS. The maximum relative errors of the voltage magnitudes under the normal scheduling and upward-reserve-activated scenarios are smaller compared to those under the downward-reserve-activated scenario.

*4.4 Computational performance of A-RBRD*

Table 3 shows the computation performance of the RBRD algorithm and the proposed A-RBRD algorithm under three different cases. We observe that the proposed A-RBRD algorithm can obtain identical objectives to the RBRD algorithm with the same number of iterations, verifying the accuracy of A-RBRD. However, the computation time of the proposed A-RBRD algorithm is significantly reduced compared to the RBRD algorithm, which demonstrates the effectiveness of the proposed A-RBRD algorithm. Table 4 lists the number of integer variables of the master problem at each iteration of the RBRD and A-RBRD algorithms. We observe that the A-RBRD algorithm significantly reduces the number of integer variables added to the master problem at each iteration.

**Table 3:** Computational performance of two algorithms

| | Algorithm | Objective | Iteration | Time (s) |
|---|---|---|---|---|
| **Case 2** | RBRD | 35981.592 | 3 | 4073.741 |
| | A-RBRD | 35981.592 | 3 | 1797.647 |
| **Case 3** | RBRD | 35180.472 | 3 | 9240.640 |
| | A-RBRD | 35180.472 | 3 | 3493.140 |
| **Case 5** | RBRD | 194568.710 | 2 | 10916.566 |
| | A-RBRD | 194568.710 | 2 | 4527.268 |

**Table 4:** Number of integer variables of master problem at each iteration of two algorithms

| | Algorithm | Number of integer variables of **MP** | | |
|---|---|---|---|---|
| | | Iteration 1 | Iteration 2 | Iteration 3 |
| **Case 2** | RBRD | 3360 | 18624 | 19048 |
| | A-RBRD | 3360 | 5858 | 5893 |
| **Case 3** | RBRD | 3360 | 18624 | 19048 |
| | A-RBRD | 3360 | 5908 | 5943 |
| **Case 5** | RBRD | 13416 | 49992 | / |
| | A-RBRD | 13416 | 20303 | / |

# 5 Conclusion

This article proposes a single-leader-multi-follower Stackelberg game model for a DER aggregator with all DERs in a distribution system to participate in day-ahead energy and reserve markets, considering the security check of the scheduling schemes of DERs. The security check problem under three different scenarios operated by the DSO is involved in the follower level. A linearized power flow model of the distribution system considering substation transformer tap is established to simplify the security check problem. Then the proposed Stackelberg game model is reformulated as a BMILP problem with only MILP follower-level problems based on the KKT-based reformulation approach. Next, an accelerated relaxation-based bi-level reformulation and decomposition algorithm is proposed to solve the BMILP problem. The simulation results for a constructed integrated T&D system and a practical integrated T&D system verify the effectiveness of the proposed Stackelberg game model. The comparisons between the cases with and without security check demonstrate that the constraints of distribution system for secure operation decrease the available downward reserve and total profits of the DER aggregator. In addition, the effectiveness of the proposed accelerated method has been proven.

# 6 Acknowledgment

This work was supported by National Natural Science Foundation of China under grant no. 52077083.

# 7 Appendix

**Appendix A: Linearization of power flow model with substation transformer in the DS**

The operation of the DS in steady state can be described as follows:

$$\begin{bmatrix} I_n \\ \boldsymbol{I} \end{bmatrix} = \begin{bmatrix} y_{st} & \tilde{\boldsymbol{Y}} \\ \tilde{\boldsymbol{Y}}^T & \boldsymbol{Y} \end{bmatrix} \begin{bmatrix} V_n \mathrm{e}^{\mathrm{j}\theta_n} \\ \boldsymbol{V} \end{bmatrix} = \begin{bmatrix} y_{st} & \tilde{\boldsymbol{Y}} \\ \tilde{\boldsymbol{Y}}^T & \boldsymbol{Y} \end{bmatrix} \begin{bmatrix} kV_{dr} \mathrm{e}^{\mathrm{j}\theta_{dr}} \\ \boldsymbol{V} \end{bmatrix}. \tag{A-1}$$

The complex-power can be expressed as follows:

$$\boldsymbol{S} = \mathrm{diag}(\boldsymbol{V})\boldsymbol{I}^* = \mathrm{diag}(\boldsymbol{V})\left(\boldsymbol{Y}^*\boldsymbol{V}^* + \tilde{\boldsymbol{Y}}^* kV_{dr}\mathrm{e}^{-\mathrm{j}\theta_{dr}}\right). \tag{A-2}$$

By expressing $\boldsymbol{V} = \boldsymbol{V}_{nom} + \Delta\boldsymbol{V}$, where $\boldsymbol{V}_{nom}$ is the pre-defined nominal voltage, and $\Delta\boldsymbol{V}$ denotes the perturbations around $\boldsymbol{V}_{nom}$, the complex-power can be transformed as follows:

$$\begin{aligned} \boldsymbol{S} &= \mathrm{diag}\left(\boldsymbol{V}_{nom} + \Delta\boldsymbol{V}\right)\left(\boldsymbol{Y}^*\left(\boldsymbol{V}_{nom} + \Delta\boldsymbol{V}\right)^* + \tilde{\boldsymbol{Y}}^* kV_{dr}\mathrm{e}^{-\mathrm{j}\theta_{dr}}\right) \\ &= \mathrm{diag}\left(\boldsymbol{V}_{nom}\right)\boldsymbol{Y}^*\left(\boldsymbol{V}_{nom} + \Delta\boldsymbol{V}\right)^* + \mathrm{diag}\left(\boldsymbol{Y}^*\boldsymbol{V}_{nom}^*\right)\Delta\boldsymbol{V} + \mathrm{diag}(\Delta\boldsymbol{V})\boldsymbol{Y}^*\Delta\boldsymbol{V}^* + kV_{dr}\mathrm{e}^{-\mathrm{j}\theta_{dr}}\,\mathrm{diag}\left(\tilde{\boldsymbol{Y}}^*\right)\left(\boldsymbol{V}_{nom} + \Delta\boldsymbol{V}\right) \end{aligned}. \tag{A-3}$$

In this paper, we select the nominal voltage that corresponds to the voltage across the network with zero injections as follows:

$$\boldsymbol{V}_{nom} = -kV_{dr}\mathrm{e}^{j\theta_{dr}}\boldsymbol{Y}^{-1}\tilde{\boldsymbol{Y}}. \tag{A-4}$$

Because the shunt impedance of the DS is negligible, we can obtain $\boldsymbol{Y}\boldsymbol{1}_N + \tilde{\boldsymbol{Y}} = \boldsymbol{Y}_{sh} = \boldsymbol{0}$ and $\boldsymbol{V}_{nom} = -kV_{dr}\mathrm{e}^{j\theta_{dr}}\boldsymbol{1}_N$. By neglecting the second-order term $\mathrm{diag}(\Delta\boldsymbol{V})\boldsymbol{Y}^*\Delta\boldsymbol{V}^*$, we can obtain the solution for $\Delta\boldsymbol{V}$:

$$\Delta\boldsymbol{V} = \boldsymbol{Y}^{-1}\,\mathrm{diag}\left(1/\boldsymbol{V}_{nom}^*\right)\boldsymbol{S}^*. \tag{A-5}$$

Here, we consider $V_{dr} = 1$ and $\theta_{dr} = 0$. By expressing the impedance matrix as $\boldsymbol{Z} = \boldsymbol{Y}^{-1} = \boldsymbol{R} + j\boldsymbol{X}$ and the complex-power as $\boldsymbol{S} = \boldsymbol{P} + j\boldsymbol{Q}$, we can obtain the real and imaginary components of $\Delta\boldsymbol{V}$:

$$\Delta\boldsymbol{V}_{re} = \frac{1}{k}\left(\boldsymbol{RP} + \boldsymbol{XQ}\right), \tag{A-6a}$$

$$\Delta\boldsymbol{V}_{im} = \frac{1}{k}\left(\boldsymbol{XP} - \boldsymbol{RQ}\right). \tag{A-6b}$$

Consequently, the approximated power flow model in the DS considering the substation transformer tap can be described as follows:

$$kV_i \approx ke_i = k\left(V_{nom,i} + \Delta V_{re,i}\right) = k^2 + \sum_{j\in N_{DS}}\left(R_{ij}P_j + X_{ij}Q_j\right), \tag{A-7a}$$

$$k\theta_i \approx kf_i = k\Delta V_{im,i} = \sum_{j\in N_{DS}}\left(X_{ij}P_j - R_{ij}Q_j\right). \tag{A-7b}$$

The bilinear terms $kV_i$ and $k\theta_i$ and the quadratic term $k^2$ in (A-7) can be linearized based on the Big-M method [24].